\documentclass[reprint,aps,pre,twocolumn,amsmath,amssymb,longbibliography]{revtex4-1}
\usepackage{graphicx}
\usepackage{dcolumn}
\usepackage{bm}

\usepackage{hyperref}

\renewcommand{\H}[0]{\mathcal{H}}
\newcommand{\M}[0]{\mathcal{M}}
\newcommand{\N}[0]{\mathbb{N}}
\newcommand{\R}[0]{\mathbb{R}}

\begin{document}

\title{Multifractal methodology}

\author{Hadrien Salat$^{1}$}
\email{Corresponding author: hadrien.salat.14@ucl.ac.uk}
\author{Roberto Murcio$^{2}$}
\author{Elsa Arcaute$^{1}$}
\address{$^{1}$Centre for Advanced Spatial Analysis (CASA), University College London, UK\\
$^{2}$Consumer Research Data Centre, Geography, University College London, London, UK}

\date{\today}

\begin{abstract}
Various methods have been developed independently to study the multifractality of measures in many different contexts. Although they all convey the same intuitive idea of giving a ``dimension'' to sets where a quantity scales similarly within a space, they are not necessarily equivalent on a more rigorous level. This review article aims at unifying the multifractal methodology by presenting the multifractal theoretical framework and principal practical methods, namely the moment method, the histogram method, multifractal detrended fluctuation analysis (MDFA) and modulus maxima wavelet transform (MMWT), with a comparative and interpretative eye. 
\end{abstract} 

\maketitle

\section{Introduction}

Since its introduction in the mid eighties to study turbulence signals \cite{FP,CMJS,MS}, multifractal theory has found numerous applications such as financial time series \cite{MDB,BDM,MMGA,BAML,MMA,BAM}, DNA sequences \cite{PBHSSG,OBGHMPSS}, the hierarchical resistor network model \cite{ARC}, satellite and microscopic images \cite{C,C2,GZ}, land use and prices \cite{HCWX,HCWX2,WFLXZF}, street networks \cite{AVHG,MMAB}, urban growth and hierarchies \cite{CZ,CW,C3}, quantum dynamical theory \cite{FWT}, and even music \cite{MPMDASBSG}.

From a mathematical point of view, the idea of applying fractal theory to measures was hinted by Mandelbrot as soon as 1982 \cite{M}, and was theorized more in depth later, notably by Evertsz and Mandelbrot \cite{EM}, Brown, Michon and Peyriere \cite{BMP}, Olsen \cite{O}, Riedi \cite{R}, and Pesin \cite{P} in the 1990s, and by Falconer \cite{F} in the 2000s. Based on this theoretical framework, four principal multifractal methodologies have been established to solve practical problems.

The moment method was the first method to be introduced in the mid eighties \cite{FP,HJKPS,ASW}. It can still be considered the reference method in the field because of the simplicity of its implementation, adaptability to many types of data, as well as the existence of many variants to enhance its accuracy or computational efficiency. The histogram method \cite{MS,EM,AGK} on the other hand improves greatly the run time over the moment method and is less reliant on error generating techniques. However, it only works for data offering a wide variety of scaling ranges. Multifractal detrended fluctuation analysis (MDFA) \cite{KZKBHBS,K} is a generalization of detrended fluctuation analysis (DFA), which was originally created to detect long-range monofractal correlations in DNA nucleotide sequences \cite{PBHSSG,OBGHMPSS}. It is used to remove artifacts created by nonstationarities in one-dimensional time series and uses the core idea of the moment method as its mechanic. The simplicity of its implementation allows to extend it to higher dimensions \cite{GZ}. Wavelet transform modulus maxima (WTMM) is another method originally invented for time series \cite{MBA,AADMV}. Better suited for a generalization to higher dimensions than MDFA, it is unfortunately more challenging to implement as it relies on a continuous framework while the three other methods are discretized.

This diversity of practical methods as well as the variety of domains they can be applied to enlighten the depth of the multifractal formalism. It is also one of its drawbacks, since most methodologies have been developed independently so that multifractality lacks the unity present in some older fields. This article aims at reducing this drawback by presenting the main methodologies in a common intuitive and comparative framework. Its intent is to help making an informed choice of a multifractal methodology for someone willing to study real datasets.

The paper is organized as follows. Section \ref{heuristic} proposes a unified intuitive approach of the core concepts behind monofractals and multifractals. The general multifractal framework can be grasped without prior knowledge of the Hausdorff measure and the box-counting dimension, although to fully understand the details these definitions are essential. In section \ref{field}, the main mathematical methodologies as well as the four practical multifractal methodologies mentioned above are detailed, compared and applied to binomial cascades. Each methodology is explained from the ground up and can be understood on its own, keeping in mind that the concepts explained in section \ref{heuristic} help to understand how they relate to one another. The main elements of interpretation as well as the limits of multifractal analysis are discussed in section \ref{interp}.

\section{From Monofractals to Multifractals}\label{heuristic}

The purpose of this section is to give a brief heuristic approach of monofractals, simply referred to as fractals, and of multifractals. All subsequent practical definitions and methodologies, as unrelated as they may appear at first glance, are only different interpretations of the core concepts explained here.

\subsection{Monofractals: Characterizing space}\label{monofrac}

The use of the word \emph{fractal} in various overlapping yet different contexts makes it quite confusing for someone new to fractality. The root of all its meanings was planted by Benoit Mandelbrot who coined it from the Latin word ``fractus'' which means ``broken'', as in ``too irregular to fit into classical geometry'' \cite{M}. Over the years, some have restricted its use to sets which present self-similarity or to subsets whose dimension is intuitively a fraction of the integer dimension of the set they are embedded in. The reason why most of the focus has turned towards the former type of sets is that self-similarity makes most subtly different definitions equivalent and provides effective computational tricks for practical uses.

Informally speaking, given an irregular subset $D$ of a space $A$ whose properties are well known, the fractal analyst is usually interested in either quantifying how much of the set $A$ is filled by the subset, or measuring the complexity of $D$ through the scale invariance of its details. Both goals are achieved simultaneously by choosing a well adapted definition of \emph{fractal dimension} and a method to compute it. In most practical situations, $A$ will be in fact $\R^n$ and the chosen dimension will be the box-counting dimension. Meanwhile, the mathematician may be more interested in the Hausdorff dimension, the canonical measure of local size. Other, more rarely seen, definitions include the correlation dimension for sets of random points \cite{GP} and the packing dimension, a dual to the Hausdorff dimension \cite{T}.

For most rigorously self-similar subsets encountered, Hausdorff and box-counting dimensions are in fact the same thing. Finding how much of the set $A$ is filled by the fractal subset $D$ is the same as finding by how much one needs to grow a sub-element of the figure to find the whole figure again. This is done through the relation 
\begin{equation}\label{selfsim}
\dim_\H= -\frac{\log(\text{number of copies})}{\log(\text{scaling factor})},
\end{equation}
where $\dim_\H$ is the Hausdorff dimension. Because of this, self-similarity is often treated as a synonym of fractality in the literature. By extension, the word ``fractal'' is also used for phenomena described by \emph{self-similar} functions, i.e. functions $f:D\subset\R^n\rightarrow\R$ for which there exists an $\alpha$ such that
\begin{equation}
\forall \lambda\in\R,x\in D, \ f(\lambda x)=\lambda^\alpha f(x),
\end{equation}
and in particular power-laws \cite{IV}, which are, in a sense, representations of scale invariance within the space $D$.

On the other hand, any dense subset made of a countable number of points is of dimension $0$ for the Hausdorff dimension and of dimension equal to its closure for the box-counting dimension. For example $\mathbb{Q}\cap[0,1]$ in $\R$ is of dimension $0$ for the Hausdorff dimension and of dimension $1$ for the box-counting dimension. Those definitions are therefore far from equivalent in all generality. A detailed discussion on what elements are desirable to define a suitable fractal dimension can be found in chapter 3 of \cite{F}.

\begin{figure}
      \includegraphics[clip,width=0.7\columnwidth,keepaspectratio]{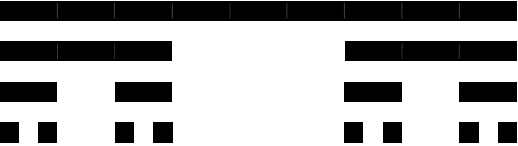}
			\caption{\textbf{Fractal middle third Cantor set.} From an initial segment of length 1, two sub-segments of length one third are created, and so on for each new segment, generating a self-similar fractal of dimension $\log(2)/\log(3)$.
			\label{FractalCantor}}
\end{figure}

A simple example is given in Fig.~\ref{FractalCantor}. The middle third Cantor set is created from an initial segment of length 1, from which two sub-segments of length one third are extracted. This process is then repeated for each new segment, and so on. Since the resulting set is self-similar with two new copies of itself each scaled at a ratio of $1/3$, one would get from equation (\ref{selfsim}) a Hausdorff dimension of $\log(2)/\log(3)$. Calculating directly the Hausdorff dimension without using equation (\ref{selfsim}) is more involving than one would expect even for such a simple set, hence the motivation to restrict the notion of fractals to self-similar sets. The value $\log(2)/\log(3)$ represents how much of the initial segment is still present in the Cantor set after an infinite number of iterations of the generating process.

\subsection{Multifractals: A theory of measures}\label{multifrac}

While monofractals are mostly concerned with spaces, multifractals deal with measures. Even if the idea behind multifractals is also to study the complexity and reveal the scaling properties of a mathematical object, those two concepts are distinct. Indeed, a measure can be a multifractal despite its support not being a monofractal \cite{SM}. Let us consider a subset $D\subset\R^n$ on which are defined:
\begin{itemize}
	\item a ``fractal'' measurement method $\M$;
	\item a finite measure $\mu$ which we want to study.
\end{itemize}
Here, $\M$ can be any method providing a way to compute a monofractal dimension, such as those quoted in the previous section, as deemed appropriate for the nature of the space $D$.

A \emph{multifractal measure} $\mu$ on $D$ is characterized by a distribution such that around any $x\in D$, the measure in a ball of radius $r$ around $x$ scales with $r$, i.e. is proportional to $r^\alpha$ for some $\alpha$, provided $r$ is small enough, and such that the sets formed by all points around which the scaling exponent is the same are monofractals for $\M$. The fractal dimension of the set corresponding to the local exponent $\alpha$ is usually denoted $f(\alpha)$.

Most methods $\M$ are based on defining a self-similar local measurement $\M_r$, such as the number of boxes of radius $r$ necessary to cover the set for the box-counting dimension or the quantity $\H_r^s$ in the definition of the $s$-dimensional Hausdorff measure (see \cite{F}). In that case, the multifractality of $\mu$ is equivalently characterized by a distribution such that the two following fundamental scaling relations hold for $r$ small enough: 
\begin{enumerate}
	\item $\mu_r(x) \sim r^{\alpha_x}$ for an $\alpha_x$ around any $x\in D$, where $\mu_r(x)$ is the measure in a ball of radius $r$ around $x$;
	\item $\M_r(\alpha) \sim r^{- f(\alpha)}$ for an $f(\alpha)$, where $\M_r(\alpha)$ is the $\M_r$-measurement of the set $\left\{x, \alpha_x=\alpha\right\}$.
\end{enumerate}

The \emph{multifractal spectrum} is the curve $f(\alpha)$ against $\alpha$. It gives, roughly speaking, the ``fractal dimension'' $f(\alpha)$ of sets where the measure scales locally with the same exponent $\alpha$. \emph{Multifractal analysis} should be understood as a method to characterize and compare measures defined on $D$ when they present enough scaling properties to alleviate the intrinsic complexity of $\left(D,\mu\right)$.

\begin{figure}[htp]
      \centering
      \includegraphics[clip,width=0.7\columnwidth,keepaspectratio]{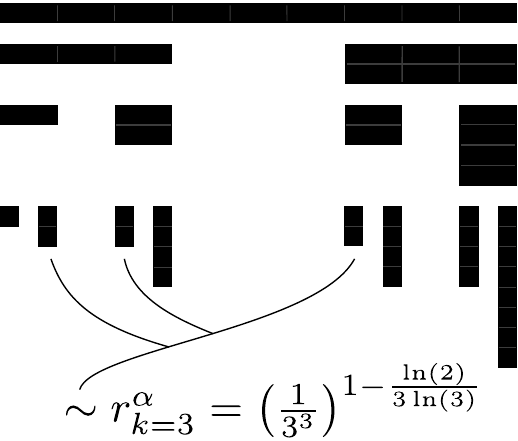}\\
			\includegraphics[clip,width=0.9\columnwidth,keepaspectratio]{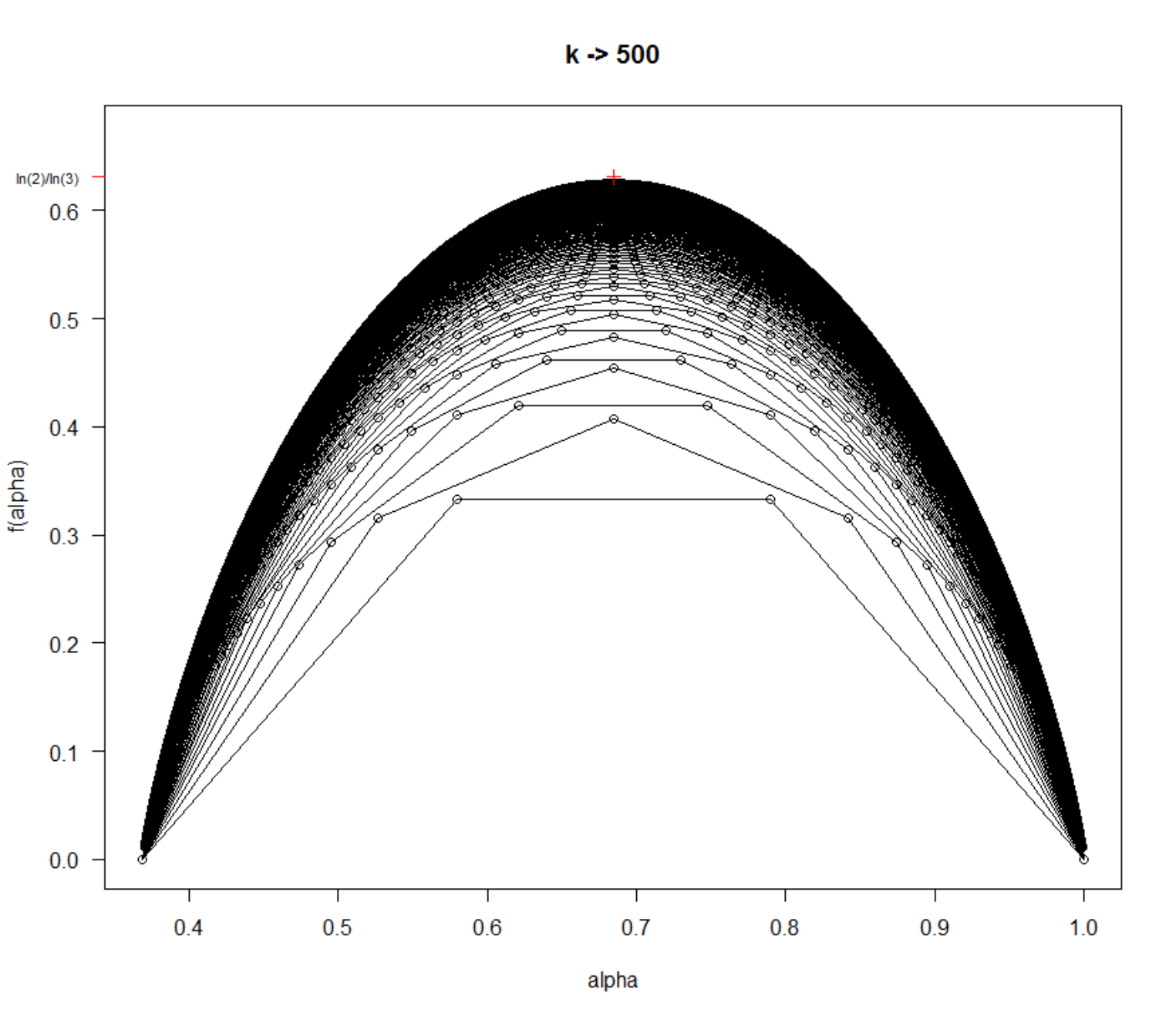}
			\caption{\textbf{Multifractal middle third Cantor set.} On the top, the first three iterations of the generation of the weighted multifractal Cantor set are represented, while on the bottom the first 500 spectra corresponding to each successive iteration for $\M$ are plotted.}
			\label{MultCantor}
\end{figure}

An example is given in Fig.~\ref{MultCantor}. The middle third Cantor set is made multifractal by weighting every right sub-interval twice as much as every left sub-interval, the total weight being normalized to 1 at each step. The first three steps of this process are illustrated in the top figure. Denote by $r_k$ the size of the new sub-intervals at step $k$, and fix $r_0=1$. Then, at step $k=3$, height sub-intervals are obtained, each of size $r_3=(1/3)^3$ and carrying a weight that can be expressed as $r_3^\alpha$ for some $\alpha$. At this macroscopic state, a broad $\M$ can be defined such that $\M_{r_k}(\alpha)$ denotes the number of sub-intervals scaling with $r_k$ for an exponent $\alpha$. This number can be in turn expressed as $r_k^{-f(\alpha)}$. For the particular $\alpha$ chosen in Fig.~\ref{MultCantor}, that is $\alpha=1-\frac{\log(2)}{3\log(3)}$, there are $3=(1/3^3)^{-1/3}$ sub-intervals carrying this measure, hence $f(\alpha)=1/3$. By repeating this calculation for each of the four different weights carried by the sub-intervals at step $k=3$, the spectrum corresponding to the bottom line of the bottom figure is obtained.

Of course, at such a low level of iteration, $\M$ does not make much sense. But as $k$ grows to infinity, the spectrum resulting from this $\M$ converges to the actual spectrum one would obtain for the Hausdorff measure and the proper totally disconnected weighted middle third Cantor set. The first 500 iterations are illustrated in Fig.~\ref{MultCantor}. It can be noted that the multifractality comes from the measure created by the weights, not from the physical support itself which is only the monofractal Cantor set presented in the previous section. In particular, the dimension of the support, here $\log(2)/\log(3)$, can be found at the peek of the spectrum.

\section{Multifractals in the Field}\label{field}

In this section, different ways of defining the ``multifractal spectrum'' are given, along with methods to compute it wherever possible. Those are all equivalent definitions in the sense that they convey the same intuitive idea of providing a fractal dimension to iso-scaling sets in data. As such, it is expected that the spectra resulting from each of them will share the same symmetries. They are not equivalent however on a more precise level, and the resulting spectra may differ in width or lead to an overshooting of the $f(\alpha)$ value. The first formal definition should be considered as canonical and subsequent definitions may be considered as approximations of it.

Methods are sorted in three groups depending on the type of data best suited to support the studied measure: mathematical abstract sets, two-dimensional data such as maps and images, and one-dimensional time series. Nonetheless, all these methods can be extended quite easily to any subset of $\R^n$. Methods from the two latter groups are tested against two samples obtained from binomial cascades. The first sample corresponds to a theoretical binomial cascade of parameter $p=0.6$, that is the measure one would obtain after averaging over an infinite number of random walks through the cascade, while the second one corresponds to one particular realization of a random walk through the cascade.

\subsection{Abstract sets - Formal definitions}\label{mathdef}

Falconer distinguishes two variants of spectra of particular interest for mathematicians \cite{F}: The \emph{singularity spectrum}, which is the most canonical definition and encompasses the universality sought in mathematics, and the \emph{coarse spectrum}, which is more adequate for practical purposes.

Consider a topological space $D$ and a finite measure $\mu$ on $D$. The local scaling exponent $\alpha_x$ of $\mu$ at $x\in D$ is given by the H\"{o}lder dimension $\dim_{loc}$, defined by 
\begin{equation}
\dim_{loc}\mu(x):=\lim_{r\rightarrow  0} \frac{\log \mu\left(B(x,r)\right)}{\log r},
\end{equation}
where $B(x,r)$ is the ball of center $x$ and radius $r$ for the topology of $D$. The \emph{singularity spectrum} is then defined by the function
$$f_\H(\alpha):=\dim_\H \left\{x\in D, \ \dim_{loc}\mu(x)=\alpha\right\},$$ where $\dim_\H$ is the Hausdorff dimension.

Note that the Hausdorff dimension is chosen for $\M$ instead of box-counting since $\left\{x\in D, \ \dim_{loc}\mu(x)=\alpha\right\}$ is often dense in the support of $\mu$, in which case box-counting would give a constant spectrum equal to the dimension of the support of $\mu$.

Let us now consider an $r$-mesh grid covering $D$ and count the number of cells for which $\mu$ is roughly $r^\alpha$. Define,
\begin{equation}
N_r(\alpha):=\#\left\{r\text{-mesh cubes} \ C, \ \mu(C)\geq r^\alpha\right\},
\end{equation}
where $\#$ stands for ``the number of''. Provided the limits do exist, the \emph{coarse spectrum} is defined by the function
$$f_C(\alpha):=\lim_{\varepsilon\rightarrow 0}\lim_{r\rightarrow 0}\frac{\log^+\left(N_r(\alpha+\varepsilon)-N_r(\alpha-\varepsilon)\right)}{-\log r},$$ where $\log^+(\cdot)$ stands for $\max(\log(\cdot),0)$.

When $f_C$ does exist, then for all $\alpha$,
\begin{equation}
f_\H(\alpha)\leq f_C(\alpha),
\end{equation}
and the equality holds true for self-similar measures (Proposition 17.9 of \cite{F}). When $f_C$ does not exist, one can define the \emph{lower} and \emph{upper} spectra by
$$\underline{f}_C(\alpha):=\lim_{\varepsilon\rightarrow 0}\liminf_{r\rightarrow 0}\frac{\log^+\left(N_r(\alpha+\varepsilon)-N_r(\alpha-\varepsilon)\right)}{-\log r},$$
and
$$\bar{f}_C(\alpha):=\lim_{\varepsilon\rightarrow 0}\limsup_{r\rightarrow 0}\frac{\log^+\left(N_r(\alpha+\varepsilon)-N_r(\alpha-\varepsilon)\right)}{-\log r}.$$
In that case, according to lemma 17.3 of \cite{F},
\begin{equation}
f_\H(\alpha)\leq\underline{f}_C(\alpha)\leq\bar{f}_C(\alpha).
\end{equation}

\begin{figure}[htp]
      \includegraphics[clip,width=0.8\columnwidth,keepaspectratio]{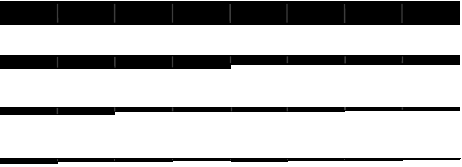}\\
			\includegraphics[clip,width=0.9\columnwidth,keepaspectratio]{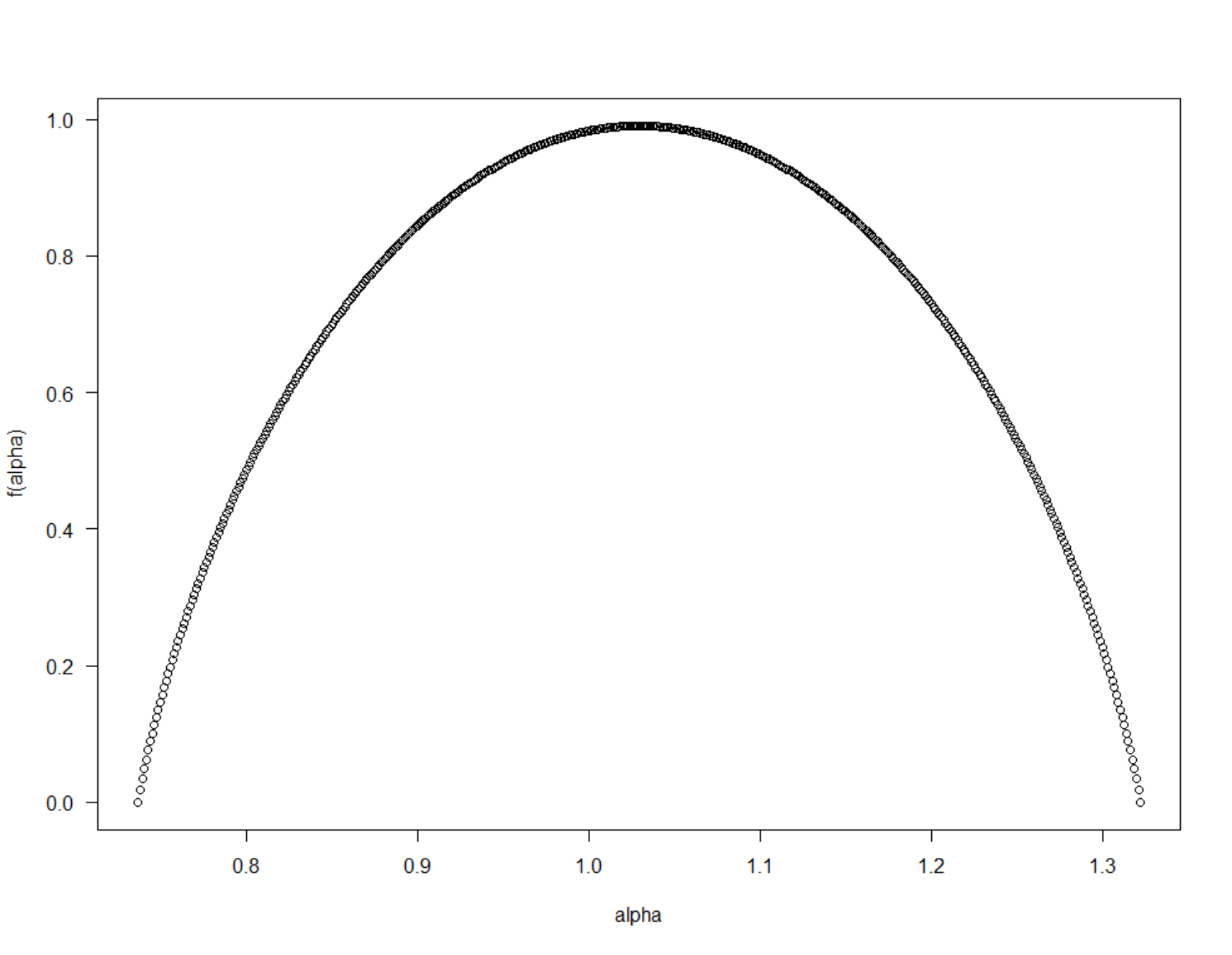}
			\caption{\textbf{Multifractal binomial cascade.} On the top, an original interval of size $1$ is divided into two sub-intervals of length $1/2$ carrying a probability $0.6$ for the left one and $0.4$ for the right one. On the bottom, the corresponding multifractal spectrum.}
			\label{BinomCascade}
\end{figure}

An example is given in Fig.~\ref{BinomCascade} for a binomial cascade of parameter $p=0.6$. The binomial cascade is a simpler version of the multifractal middle-third Cantor set introduced in section \ref{multifrac}. Here, an original interval of size $1$ is divided into two sub-intervals of length $1/2$ carrying a probability $0.6$ for the left one and $0.4$ for the right one. This process is then iterated on each resulting sub-intervals, and so on. The resulting singularity spectrum $f_\H$ in the bottom figure is computed using the same trick as in section \ref{multifrac} and is identical to the coarse spectrum $f_C=\underline{f}_C(\alpha)=\bar{f}_C(\alpha)$, since the measure is self-similar. Here, the fractal dimension of the support is 1 since the iterative process does not create ``holes'' in the initial segment.

\subsection{Spatial data - Moment and histogram methods}\label{SpatData}

Moment and histogram methods are the elementary components of practical multifractal spatial data and image analysis. Both methods rely on counting the measure at different levels of aggregation and include a multitude of variants depending on the aggregation method chosen. The most basic way to aggregate consists in applying square grids of increasing resolutions to the data. Instead of grids, one can use ball neighborhoods of increasing radius as they can be easily calculated for two-dimensional geographic data by GIS software, or gliding boxes to increase the number of data points. Grids can also be based on any regular unit shape, such as diamonds or equilateral triangles, instead of squares, to enhance computational complexity or suitability with the data. Other variants exist to counter the reliance on error generating techniques such as linear fits or Legendre transforms. The core mechanics of both methods will be detailed and reference to the literature will be given for their variants.

Let us consider a mesh grid of unit $r$ covering a domain $D$, and a phenomenon occurring $N$ times in $D$. 
\begin{equation}
p_i:=N_i/N=\int_{i^{th} box} d\mu(x)
\end{equation}
is the probability that an instance of the phenomenon occurs in the $i^{th}$ box. To interpret the two scaling rules from section \ref{multifrac}, one simply needs
\begin{enumerate}
	\item $p_i\sim r^{\alpha_i}$;
	\item $N(\alpha_i) \sim \rho(\alpha_i)d\alpha_i r^{-f(\alpha_i)}$,
\end{enumerate}
where $N(\alpha_i)$ is the number of times $\alpha$ falls in each interval $[\alpha_i,\alpha_i+d\alpha_i]$, and $\rho$ is a density function used to take into account the dimension of $D$.

To effectively compute $f$, the trick generally used is the \emph{moment method} \cite{FP,HJKPS,ASW}. By raising $p_i$ to its moment $p_i^q$ for different $q$, one can force only one value of alpha for each $q$ to make a significant contribution to the total value of the measure. Consider
\begin{equation}\label{BaseMoment}
\begin{split}
Z(q):=\sum_i p_i^q &\sim \sum_i r^{\alpha_i q}\\
                   &\sim \int_\alpha N(\alpha)r^{\alpha q}\\
									 &\sim \int\rho(\alpha)r^{\alpha q-f(\alpha)}d\alpha,
\end{split}
\end{equation}
then, for $r$ small enough, the value of $Z(q)$ is almost entirely given by the $\alpha$ such that 
\begin{equation}
\tau(q):=\alpha q - f(\alpha)
\end{equation}
is minimal. Let us call $\alpha(q)$ this value of $\alpha$. It is easy to show by a Legendre transform that the minimality condition yields 
\begin{equation}\label{eq10}
\alpha(q)=\frac{d\tau(q)}{dq};
\end{equation} 
and 
\begin{equation}\label{eq11}
f(\alpha(q))=\alpha(q)q-\tau(q),
\end{equation}
so that computing $\tau(q)$ from $Z(q)$ for each $q$ between $-\infty$ and $+\infty$ is enough to obtain the full spectrum.

It is not possible in practice to use an infinite range of values for $q$, nor is it desirable since the method becomes less and less accurate for extreme values of $q$. To select an appropriate range of $q$, one should set a threshold for the error generated by linear fitting and dismiss all the values of $q$ for which the threshold is exceeded. It is also necessary to select $q$ in order to ensure that $f(\alpha)>0$ and that the \emph{generalized dimension} $D_q$, defined by $D_q:=\tau(q)/(q-1)$, remains lower than the dimension of the physical support of the phenomenon. The reason for that last constraint will become clear when the physical meaning of $D_q$ will be explained in section \ref{interp}.

In practice, $\tau(q)$ is found directly as the slope in a log-log plot of $\sum_i \mu_i^q(r)$ versus $r$ obtained for different grid sizes $r$, where $\mu_i(r)$ is the total measure of cell $i$ of size $r$. Since this slope is independent of the normalization of the measure $\mu$, $\mu$ does not need to be weighted as a probability measure. Explicitly, $\tau(q)$ is found as the limit
\begin{equation}\label{tauq}
\tau(q)=\lim_{r\rightarrow0} \frac{\log(\sum_i \mu_i^q(r))}{\log(r)}.
\end{equation}

In accordance with the idea that practical methods tend to create overshooting, one finds that
\begin{equation}
f_\H(\alpha)\leq\underline{f}_C(\alpha)\leq\bar{f}_C(\alpha)\leq f_M(\alpha),
\end{equation}
where $f_M$ is the spectrum resulting from the moment method (corollary from Proposition 17.2 of \cite{F}).

Both finding $\tau(q)$ through linear fitting and applying numerical Legendre transforms have a cost on the accuracy of the results. The possibility of averaging over several samples can be extremely beneficial. There are two ways of doing this: averaging over a range of $\{f_j(\alpha_i)\}_j$ computed independently for different samples $j$, or averaging first over a range of $\{N_j(\alpha_i)\}_j$ and then deducing the corresponding $f(\alpha_i)$ from the relation $N(\alpha) \sim \rho(\alpha)d\alpha r^{-f(\alpha)}$ on the averaged values. The first solution guarantees to obtain a ``classic'' positive spectrum, but it can be unreliable if the fluctuations between the $f_j(\alpha_i)$ are too important. The second solution is more reliable, but may create an artificial negative part in the spectrum if $N(\alpha_i)$ falls below 1 for some $\alpha_i$ as a result of the averaging process.

Chhabra and Sreenivasan argue in \cite{CS} that this artificial negative part can still be of relevance when a strong underlying probabilistic process is suspected either as a cause of the phenomenon or as a result of the experimental methodology since it could describe the rarely occurring events. Unfortunately, since $\alpha\mapsto N(\alpha)$ decreases exponentially compared to $\alpha\mapsto f(\alpha)$ in the negative part, one would need an exponentially increasing number of samples as the resolution gets smaller to maintain accuracy while supersampling. Paradoxically, for a constant number of samples, a better resolution would mean a less accurate result.

A multiplier method is presented in \cite{CS} to tackle this problem. The self-similarity of the measure implies the existence of an underlying scale-invariant multiplier distribution such that the $\alpha_i$ at resolution $r_k$ are only the result of $k$ composition of said multipliers. If there is no correlation in the underlying probabilistic process from resolution $r_{k-1}$ to resolution $r_{k}$ for some $k$, then one can deduce the multipliers and hence $\alpha$. In particular, if all levels of resolution are uncorrelated, one can choose $k=1$, otherwise, one should choose the smallest $k$ for which a level of resolution is uncorrelated to the previous one.

Denote $r_0$ the minimal resolution and $r_k$ the resolution chosen as described above. Then, define 
\begin{equation}
r:= r_k/r_0,
\end{equation}
and for each sample $j$ and box $i$,
\begin{equation}\label{multp1}
M_{ij}:= \mu_{ij}(r_0)/\mu_{ij}(r_k).
\end{equation}
Then, according to \cite{CS}, $\tau(q)$ and $\alpha(q)$ are given by
\begin{gather}\label{multp2}
\frac{1}{N(r)}\tau(q)+d\approx-\frac{\log\left(1/N(r)\sum_{i,j}M_{ij}^q\right)}{\log(r)};\\
\frac{1}{N(r)}\alpha(q)\approx-\frac{\sum_{i,j}M_{ij}^q\log(M_{ij})}{\sum_{i,j}M_{ij}^q\log(r)}.
\end{gather}
where $N(r)$ is the number of non zero values of $M_{ij}$ and $d$ is the dimension of the physical support $D$.

Chhabra and Sreenivasan have shown that for a binary cascade and the dissipation field of fully developped turbulance in the atmospheric surface layer, using the multiplier method allowed to expand the negative part of the spectrum and make it converge to the theoretical result more rapidly than the supersampling method, with also a gain in computational complexity.

Another way to expand the set of sample points consists in using one grid and aggregate with a gliding box for different radii of said gliding box instead of using different grid sizes \cite{C,CA}. In that case, the corrected formula reads
\begin{equation}
\frac{1}{N(r)}\tau(q)+d\approx \lim_{r\rightarrow 0}\frac{\log\left(1/N(r)\sum_{i=1}^{N(r)}\mu_i^q(r)\right)}{\log(r)},
\end{equation}
where $N(r)$ is the number of gliding boxes of size $r$ with non zero measure, $\mu_i^q(r)$ is the measure inside the $i^{th}$ gliding box, and $d$ is the dimension of the physical support $D$.

\begin{figure}[htp]
      \includegraphics[clip,width=0.95\columnwidth,keepaspectratio]{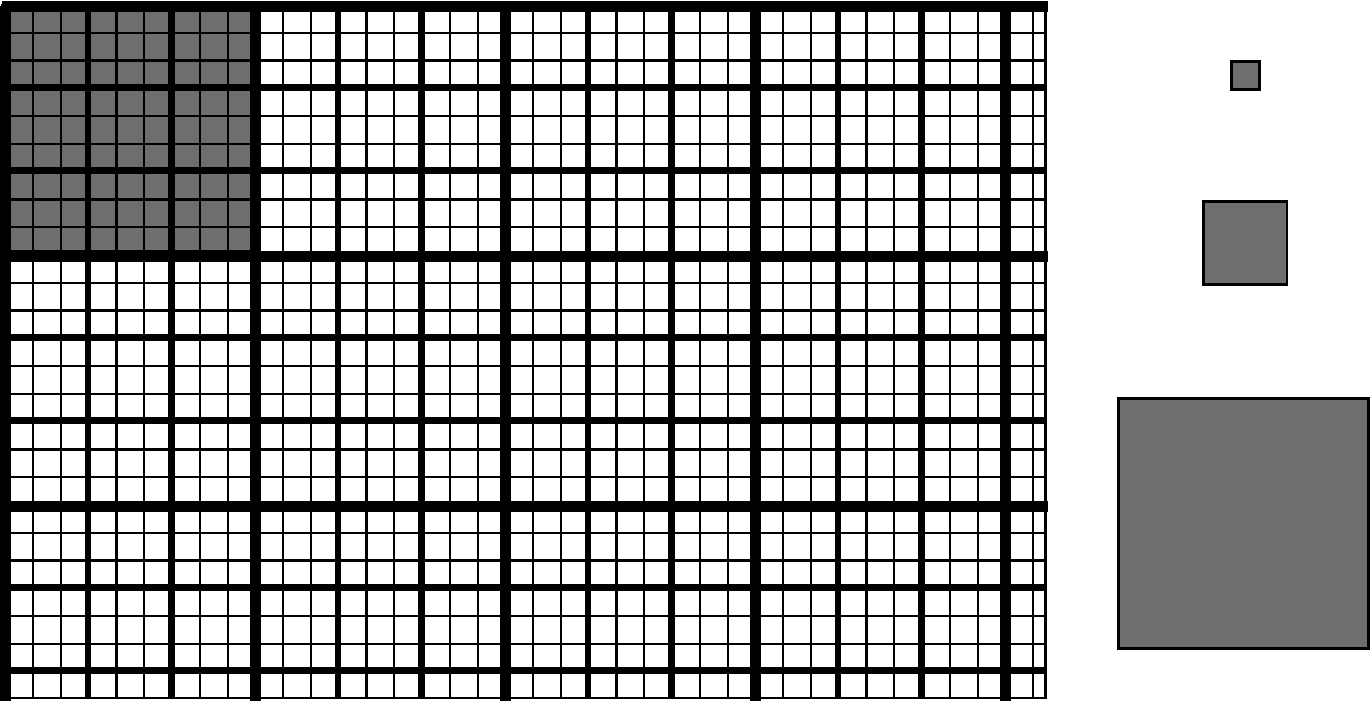}\\ \vspace{0.2cm}
			\includegraphics[clip,width=0.95\columnwidth,keepaspectratio]{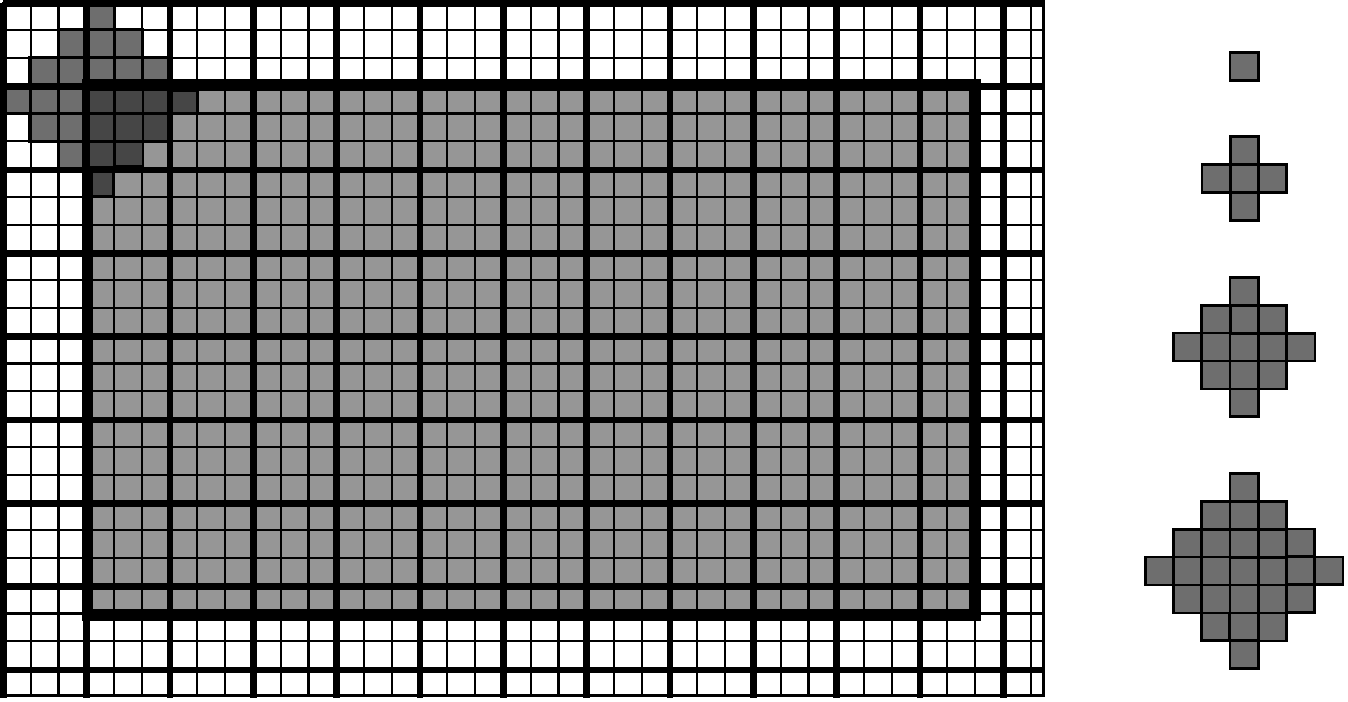}
			\caption{\textbf{Comparison between grid and gliding box upscalings.} In the top image, the third level of aggregation, which is only $3^2$ times the smallest resolution, only allows to fully place 8 boxes on the figure. In contrast, gliding boxes applied on the lowest image maintain 608 sample points at the cost of forcing to remove a wide border from the analysis.}
			\label{GBvsBC}
\end{figure}

Since gliding boxes need not be mutually exclusive, contrary to squares from a mesh grid, the number of values contributing to the analysis remains that of the smallest resolution at all scales. The trade-off is that only boxes which are completely bounded in $D$ should be included, so that only the ``inner portion'' of the data can be analyzed, or the object of study needs to be surrounded by a large neighborhood of known values (see Fig.~\ref{GBvsBC}). Using gliding boxes allows a higher raw number of sample points at the cost of restricting the range of study. It is of course possible to join gliding boxes and the multiplier method by adapting the definition of $\mu_{ij}$ and $N(r)$ in equations (\ref{multp1}) and (\ref{multp2}).

In \cite{CMJS}, Chhabra \textit{et al.} propose a recipe to avoid the Legendre transform of $\tau(q)$ when the measure arises from multiplicative processes. Once the $p_i$ have been established, compute
\begin{equation}
\mu_i^q(r)=\frac{p_i^q}{\sum_j p_j^q}.
\end{equation}
Then, the Legendre transform can be directly integrated in the calculation of $f$ and $\alpha$ through the formulas
\begin{gather}
f(q)=\lim_{r\rightarrow0}\frac{\sum\mu_i^q(r)\log\mu_i^q(r)}{\log(r)};\\
\alpha(q)=\lim_{r\rightarrow0}\frac{\sum\mu_i^q(r)\log p_i(r)}{\log(r)}.
\end{gather}
Note that here $\alpha(q)$ is the average value of $\alpha$ at resolution $q$. Unfortunately, this recipe does not remove the need for linear fitting when calculating the limits, which is usually the main cause of error. It was applied to two-scaled cantor measures in \cite{CJ} with good results when the boxes size progression matched the sub-intervals size progression, and ``satisfactory'' results otherwise despite the errors created by linear fitting. It was also found in \cite{CMJS} that the result of this direct computation were in good agreement with those obtained from Legendre transforming $\tau(q)$ for fully developed turbulence.

Another direct approach is the \emph{histogram method}, see for example \cite{MS,AGK}. The idea consists in finding the cells with extremal values of total measure for different grid resolutions, and dividing the distance between those values into regular intervals to exploit the fact that exactly one value of $\alpha$ and $f(\alpha)$ will correspond to an extremity of one of the new sub-intervals.

Let us call $\mu_i^k$ the total measure of cell $i$ of a grid of unit $r_k$, and $N(X)$ the number of boxes presenting feature $X$. Step by step, the method breaks down as follows.

\begin{enumerate}
  \item Compute $X_i^k:=\log(\mu_i^k)$ for each cell $i$ of different grids of unit $r_k$;
	\item Divide $[X_{min}^k,X_{max}^k]$ regularly in $n$ smaller intervals for each $k$, where $X_{min}^k:=\min\{X_i^k\}$ and $X_{max}^k:=\max\{X_i^k\}$;
	\item Deduce one value of $\alpha$ and $f(\alpha)$ from the slopes of $X^k$ and $N(X^k)$ versus $\log(r_k)$ for each sub-interval;
	\item Repeat for different grid positions to get a better estimate.
\end{enumerate}

In step 3, for $1\leq j\leq n$, the value $\alpha_j$ is given by the slope of $X_{j,k}$ versus $\log(r_k)$ where $X_{j,k}$ is one of the extremities of the $j^{th}$ sub-interval for grid resolution $r^k$. According to \cite{MS}, the correct normalization of the total measure leads to an expression of $f(\alpha)$ as the slope of $\log(N(X_{j,k})\Delta X^{1/2})$ versus $log(r_k)$, where $N(X_{j,k})$ is the number of boxes of size $r_k$ containing an $X$ falling in the interval of size $\Delta X$ around $X_{j,k}$.

It is indeed a problem to find the correct normalizations of $\alpha$ and $f(\alpha)$ because the relations $p_i\sim r^{\alpha_i}$ and $N(\alpha_i) \sim \rho(\alpha_i)d\alpha_i r^{-f(\alpha_i)}$ depend on prefactors that are unknown \textit{a priori}. Such a problem is absent in previous methods since those factors are canceled while taking the limit for $r\rightarrow 0$, which is not done here.

Meneveau and Sreenivasan have applied the histogram method in \cite{MS} for a binomial measure, a period doubling attractor for a specific logistic map and the dissipation field of turbulent kinetic energy in turbulence flows. They found good agreement with the results obtained from the moment method for the first two cases but it was evidenced that errors are generated by the histogram method for measures with small scaling ranges such as the third case. In fact it was evaluated that the exponent found by this method is only accurate up to order $\log(L/r)^{-2}$, where $L$ represents a characteristic value intrinsic to the problem (a translation of the unknown prefactor). It was recommended to use this method only for measures such that the largest obtainable scale is at least $10^3$ times bigger than the smallest measurable scale.

\begin{figure}[htp]
      \includegraphics[clip,width=0.9\columnwidth,keepaspectratio]{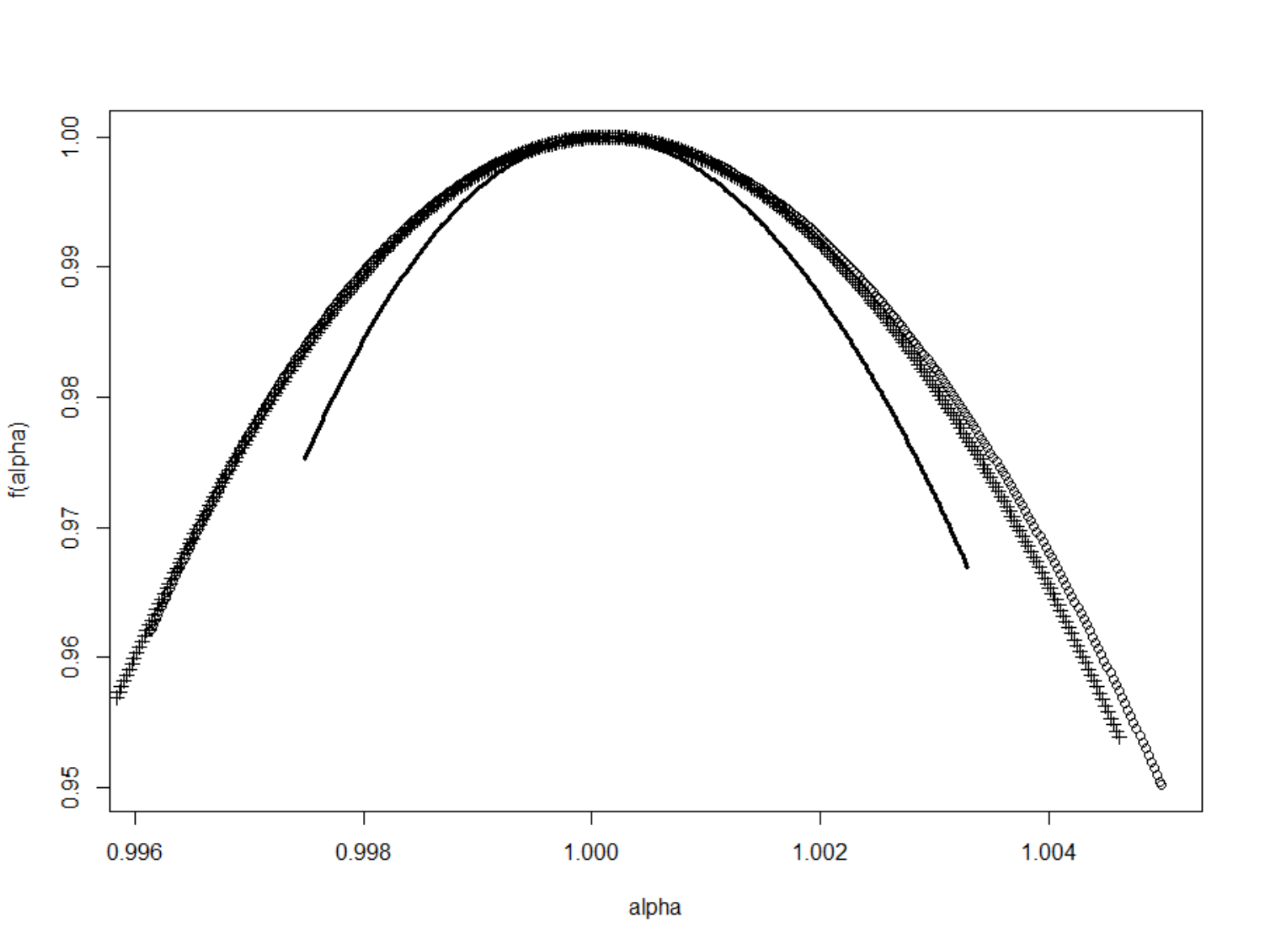}\\
			\includegraphics[clip,width=0.9\columnwidth,keepaspectratio]{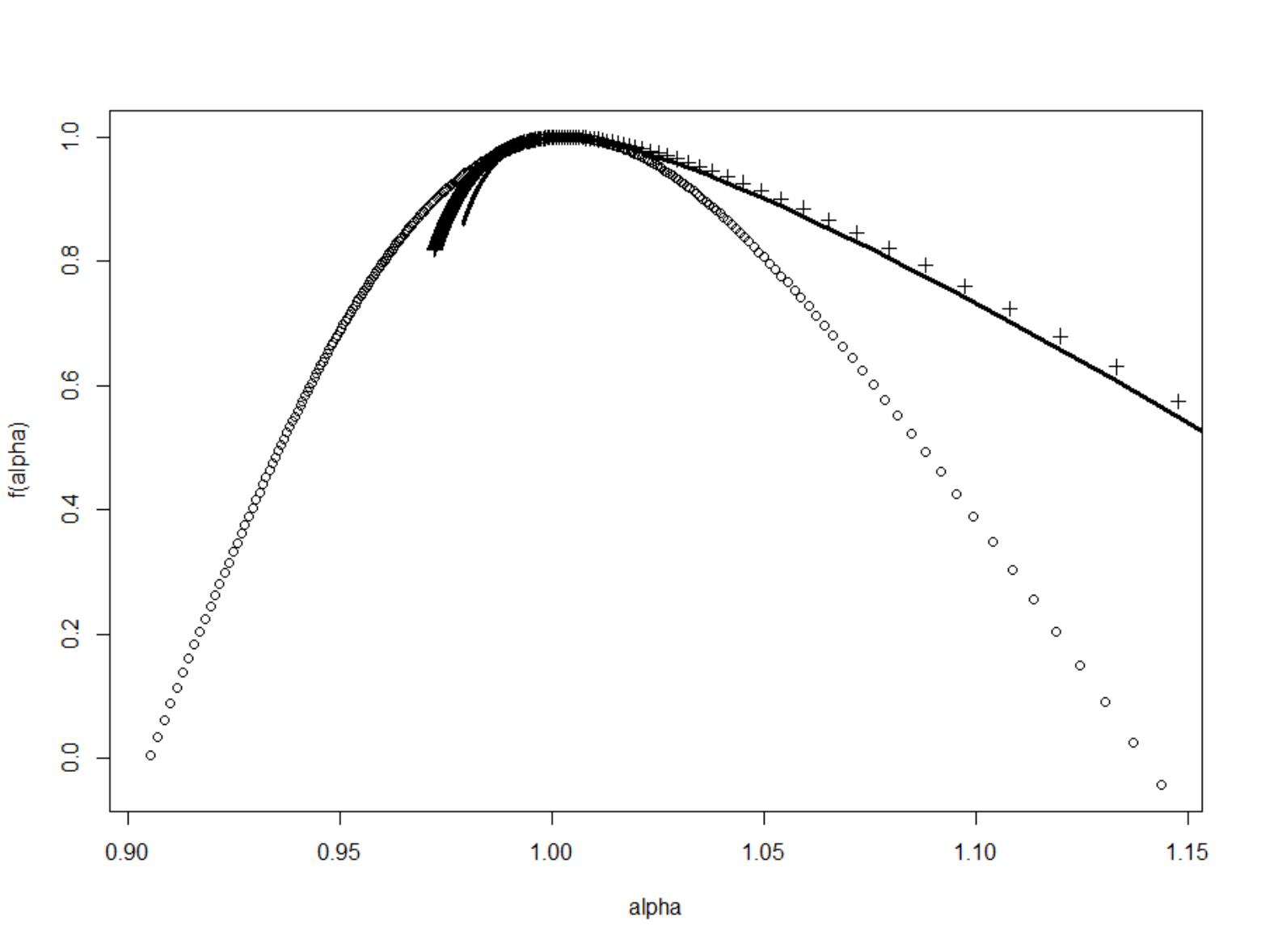}
			\caption{\textbf{Moment method and variants applied to multifractal binomial cascades.} The standard moment method (plain line), the gliding box method (circles), and the multiplier method (crosses) are applied to the theoretical cascade on the top and to a particular random walk through the cascade on the bottom.}
			\label{BinomMomentGen}
\end{figure}

\begin{figure}
      \includegraphics[clip,width=0.9\columnwidth,keepaspectratio]{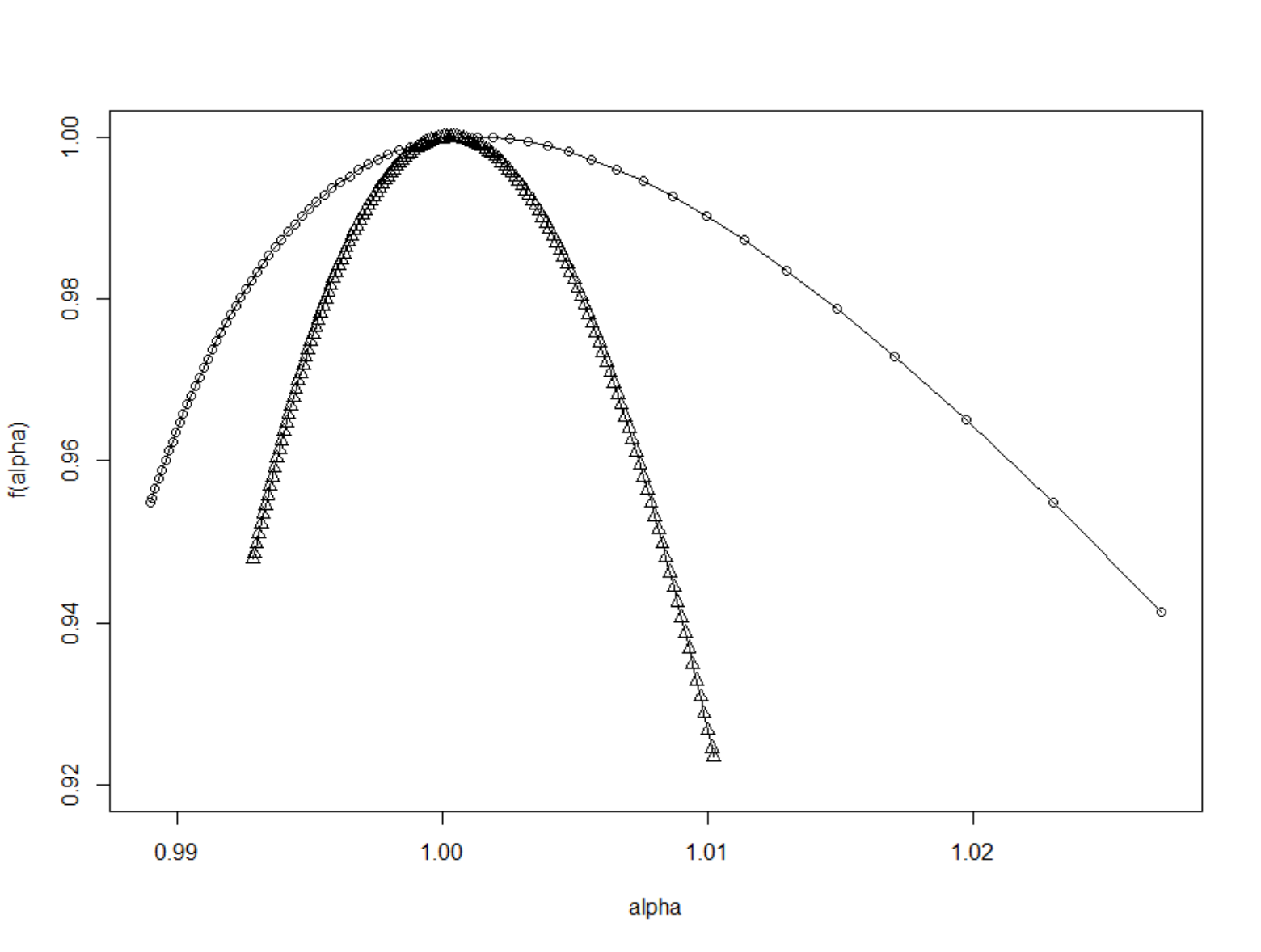}\\
			\includegraphics[clip,width=0.9\columnwidth,keepaspectratio]{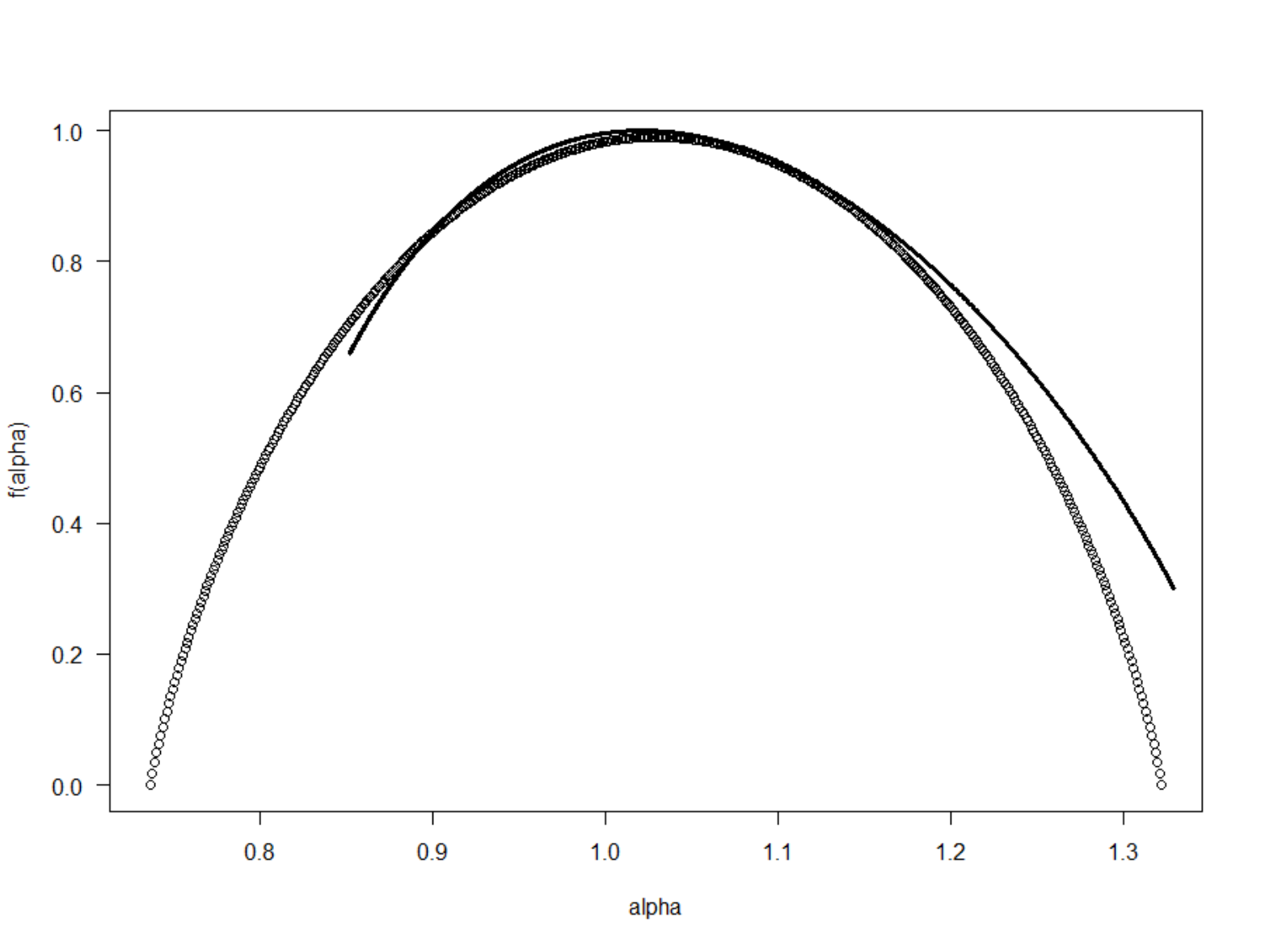}
			\caption{\textbf{Influence of the number of iterations on the spectrum.} On the top, the spectrum made of triangles is obtained for 10 repetitions of the iterative process generating the cascade, while the spectrum made of circles is obtained for 20 repetitions. Increasing the number of iterations makes the spectrum larger and therefore closer to the theoretical curve. On the bottom, a rescaling of the range of alpha on the curve resulting from the moment method (plain line) is enough to make it harmonious with the reference curve (circles).
			\label{Numberofit}}
\end{figure}

\begin{figure}[htp]
			\includegraphics[clip,width=0.9\columnwidth,keepaspectratio]{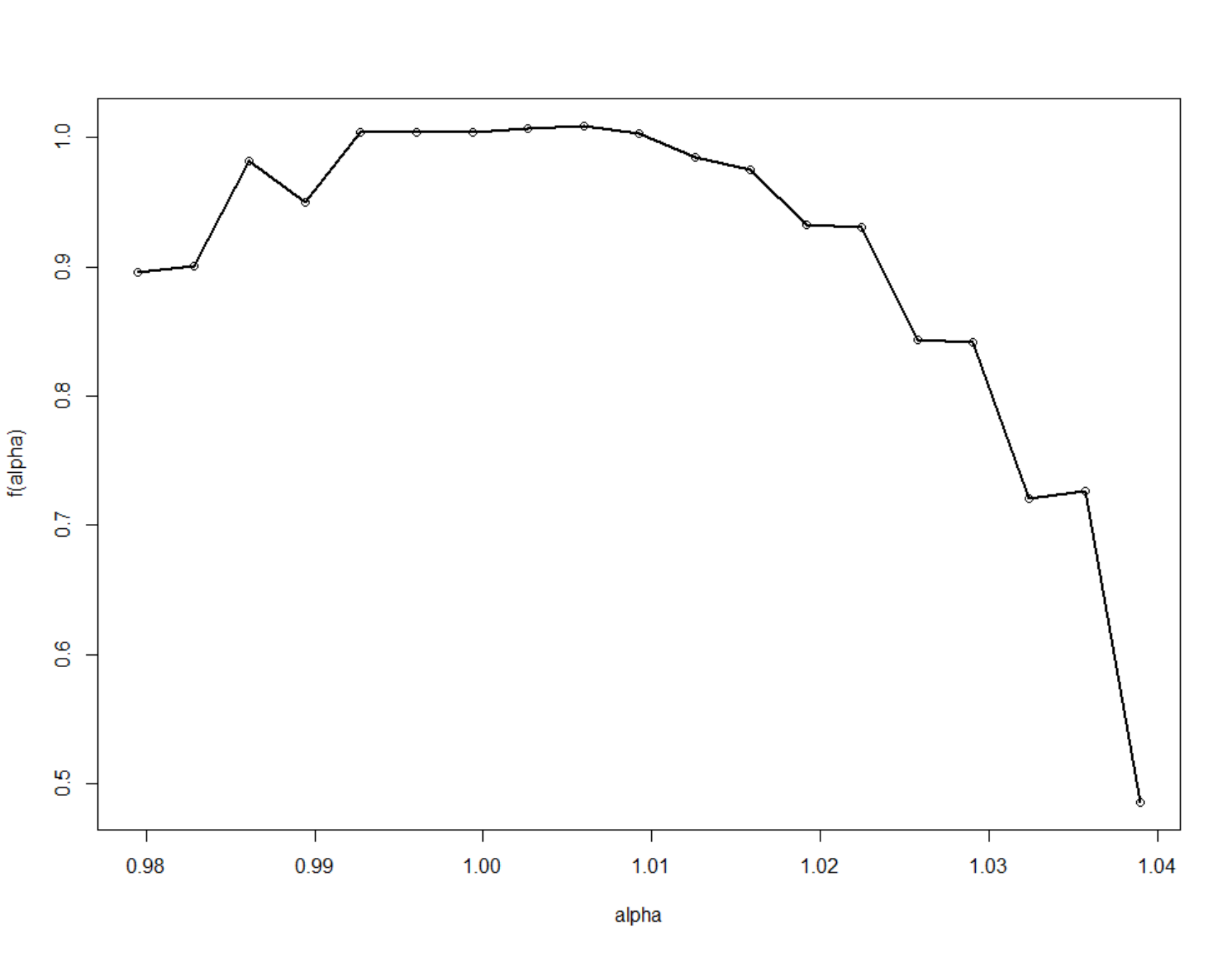}\\
			\includegraphics[clip,width=0.9\columnwidth,keepaspectratio]{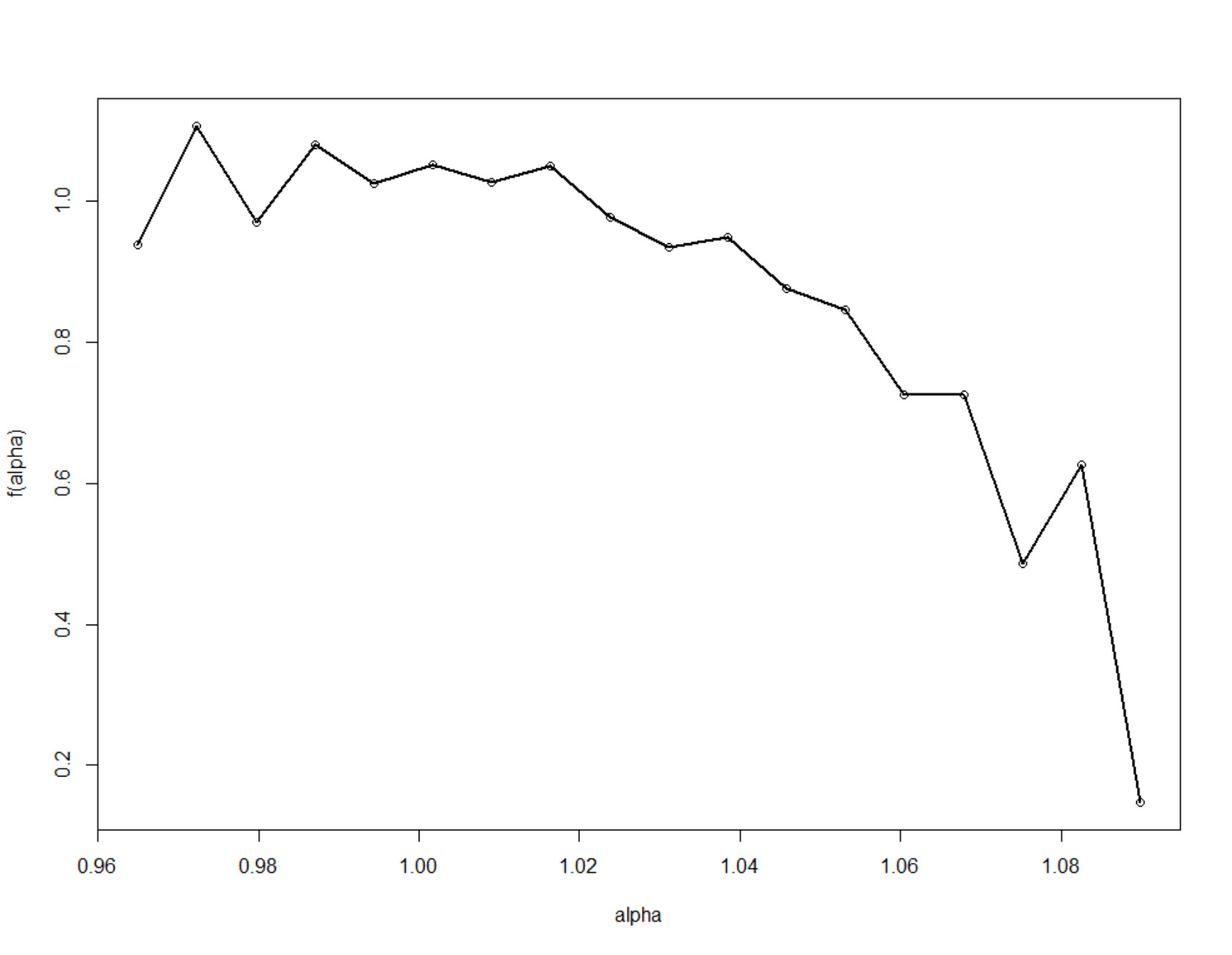}
			\caption{\textbf{Histogram method applied to binomial cascades.} On the top, a theoretical binomial cascade of parameter $p=0.6$ and on the bottom a particular realization of it.}
			\label{Histo}
\end{figure}

The spectra resulting from moment and histogram methods applied to binomial cascades are given in Fig.~\ref{BinomMomentGen}-\ref{Histo}. In Fig.~\ref{BinomMomentGen}, the plots on the top are obtained for the theoretical binomial cascade of probability $p=0.6$ introduced in section \ref{mathdef}, that is the result of averaging an infinite number of random walks through the iterative process defining the cascade. The plots on the bottom are obtained for one particular realization of a random walk through the cascade. The iterative process is stopped after the $20^{th}$ step for which over a million sub-intervals have already been created. This is done to ensure a reasonable run time and use of memory, and because it is comparable to the size of many encountered data types such as pixels of an HD image or data collected from a city-size human settlement. Unfortunately, it also means that the studied cascades are not equivalent to the one used in Fig.~\ref{BinomCascade}, for which the iterative process is repeated an infinite number of times. As such the resulting spectra are expected to be similar, but not necessarily identical to the theoretical curve of Fig.~\ref{BinomCascade} depending on the sensitivity of the chosen multifractal method. The chosen range of $q$ for the moment method goes from $-20$ to $20$.

The results of the standard moment method and its variants, the gliding box and the multiplier methods, are illustrated in Fig.~\ref{BinomMomentGen}. In the theoretical case on the top, the range of $\alpha$ is a lot narrower than what is expected based on the reference curve (Fig.~\ref{BinomCascade}). The variants help improve the situation, but not by much. This problem is due to the fact that the iterative process was stopped too soon. The top figure of Fig.~\ref{Numberofit} evidences that stopping the iterative process at step 25 results in a wider spectrum (circles) than stopping it at step 10 (triangles). Aside from this problem, the resulting spectrum is similar to the reference one and a simple rescaling of the range of $\alpha$ is enough to make both spectra harmonious.

For the particular realization of the cascade on the bottom of Fig.~\ref{BinomMomentGen}, the accuracy of the standard and gliding box moment methods deteriorates rapidly for negative $q$. The multiplier method gives the best results overall. To keep the curve above zero, the range of q had to be restricted to $q\geq-4$ for the first two variants and to $17.5\geq q\geq-10$ for the multiplier variant.

Results of the histogram method applied to the binomial cascades can be found in Fig.~\ref{Histo}. Unfortunately, for such a small range of scaling, the histogram method is not well adapted and the resulting spectra are not smooth. The error generated by the method makes results difficult to interpret in this case. It has however the advantage of being faster than the moment method and gives a range of $\alpha$ closer to the reference in this particular case.

\subsection{Time series - MDFA \& WTMM}

Some methods have been independently developed for the specific purpose of studying time series, an important object in physics. They are therefore particularly well suited for one-dimensional data, but can be extended to any dimension at the expense of computational complexity. For our purpose, time series will be defined as a one dimensional array of discrete values representing observations taken at regular intervals.

\emph{Multifractal Detrended Fluctuation Analysis} (\emph{MDFA}) is thoroughly described in \cite{KZKBHBS,K}. In the basic approach, time series are first sub-divided into smaller segments on which is subtracted a least-squares best-fit polynomial of a chosen order to remove the artifacts created by nonstationarities in the time series. A method similar to the moment method is then applied to the resulting detrended series. In details, MDFA consists of the following steps.
\begin{enumerate}
	\item Divide a time series $f$ into $N_s$ segments containing $s$ elements each for an array of $s$;
	\item For each $s$ and on each segment, replace $f(\cdot)$ with its cumulative sum $F(\cdot,s)$;
	\item Detrend by removing a least-squares fitted polynomial of order $n$ to $F$ on each cumulative segment;
	\item Denoting $\bar{F}$ the result of step 2, compute $$F_q(s):=\left(\frac{1}{N_s}\sum_{\nu=1}^{N_s}\bar{F}(\nu,s)^q\right)^{1/q};$$
	\item Find the scaling relation $F_q(s) \sim s^{h(q)}$.
\end{enumerate}

Here, $h(q)$ is the \emph{hurst exponent}, which relates to the classical $\tau(q)$ through the relation $\tau(q)=q h(q)-D_f$, where $D_f$ is the fractal dimension of the physical support of $f$.

The second step is not compulsory but is helpful in the sense that it allows the use of simple polynomials of the form $a_ni^n+\dots+a_0$ with $i\in\N$ to detrend in step three. It should be noted that the expression of $F_q$ given above is not well defined for $q=0$. It is indeed necessary to set
\begin{equation}
F_0(s)=\exp\left(\frac{1}{N_s}\sum_{\nu=1}^{N_s}\log\left(\bar{F}(\nu,s)\right)\right).
\end{equation}
According to \cite{KZKBHBS}, MDFA works only for positive $h$ and becomes inaccurate for $h$ close to $0$. A solution consists in integrating by considering the sum $\sum F(\cdot,s)$ instead of $F$. Following the same steps, one would obtain $h(q)+1$ instead of $h(q)$.

The use of $\tau(q)$ as an intermediary step is given to link the method to previous techniques (see equation (\ref{eq10}) and (\ref{eq11})), but one can compute directly $\alpha$ and $f(\alpha)$ using the expressions
\begin{gather}
\alpha(q)=h(q)+q\frac{dh(q)}{dq};\\
f(\alpha(q))=q(\alpha-h(q))+D_f.
\end{gather}

An original application of MDFA, presented in \cite{KZKBHBS}, is to distinguish the underlying cause of multifractality between long-range correlations and a broad probability density function. Indeed, if one shuffles the time series, all correlations are destroyed. Hence, when applying MDFA to the shuffled time series, if the resulting spectrum shifts towards monofractality, that is if the shuffled $h$ is constant, then multifractality is probably due to long range correlations. Obviously, multifractality can have several causes, but an important influence of correlations should be noticeable in the alteration of $h$.

MDFA can be extended to 2 or more dimensions using multivariate polynomials, as proposed in \cite{GZ}. The 3 dimensions extension is particularly useful to study the evolution of a two-dimensional spatial pattern simultaneously in space and time. Unfortunately, the necessity to choose a common array of $s$ for all directions at the same time, and therefore constraining the precision and accuracy of the method to the direction along which the data is the most scarce or irregular, as well as the rapidly growing computational complexity are significant limiting factors.

\emph{Wavelet Transform Modulus Maxima} (WTMM) replaces square intervals (and square grids for higher dimensions) by highly customizable wavelets. If those are chosen orthogonal to low order polynomials, a natural detrending happens. The ``modulus maxima'' part of the name refers to an observation that analyzing the data along maxima lines is enough to bring out the underlying multifractal structure. With a judicious choice of wavelets, one can therefore integrate the advantages of MDFA and extend it efficiently to higher dimensions while maintaining adequate computational complexity. The wavelet transform idea is introduced in \cite{MBA} and a more extensive study of WTMM can be found in \cite{AADMV}.

Consider a function $f:\R\rightarrow\R$ representing either a continuous signal or the interpolation of a time series and a wavelet $\psi$ orthogonal to low-order polynomials. The wavelet is a real valued function, preferably of zero mean to ensure the method is invertible. WTMM is divided in the following steps.
\begin{enumerate}
	\item Operate the wavelet transform by defining for any $x_0$: $$T_\psi[f](x_0,r):=\frac{1}{r}\int_{-\infty}^{+\infty}f(x)\psi\left(\frac{x-x_0}{r}\right)dx;$$
	\item Sum along the local maxima lines $\mathcal{L}(r)$ at scale $r$: $$Z_q(r)=\sum_{l\in\mathcal{L}(r)}\left(\sup_{(x,\tilde{r})\in l} \left|T_\psi[f](x,\tilde{r})\right|\right)^q;$$
	\item Find the scaling relation $Z_q(r) \sim r^{\tau(q)}$.
\end{enumerate}

The set of maxima lines $\mathcal{L}(r)$ is defined as follows. Consider the set of extrema $L(r)$ defined by
\begin{equation}
L(r):=\left\{x, \ \frac{\partial}{\partial x}\left(x\mapsto\left|T_\psi\left[f\right](x,r)\right|\right)=0\right\}.
\end{equation}
Then, the set $\{(x,r), \ x\in L(r)\}$ is formed of connected curves called \emph{maxima lines}. The set $\mathcal{L}(r)$ is then obtained as the set of all maxima lines defined for all $r'\leq r$. Explicitly,
\begin{equation}
\mathcal{L}(r):=\left\{\left(x(r'),r'\right), \ \forall 0\leq r'\leq r, x(r')\in L(r')\right\}.
\end{equation}

\begin{figure}
      \includegraphics[width=\columnwidth,keepaspectratio]{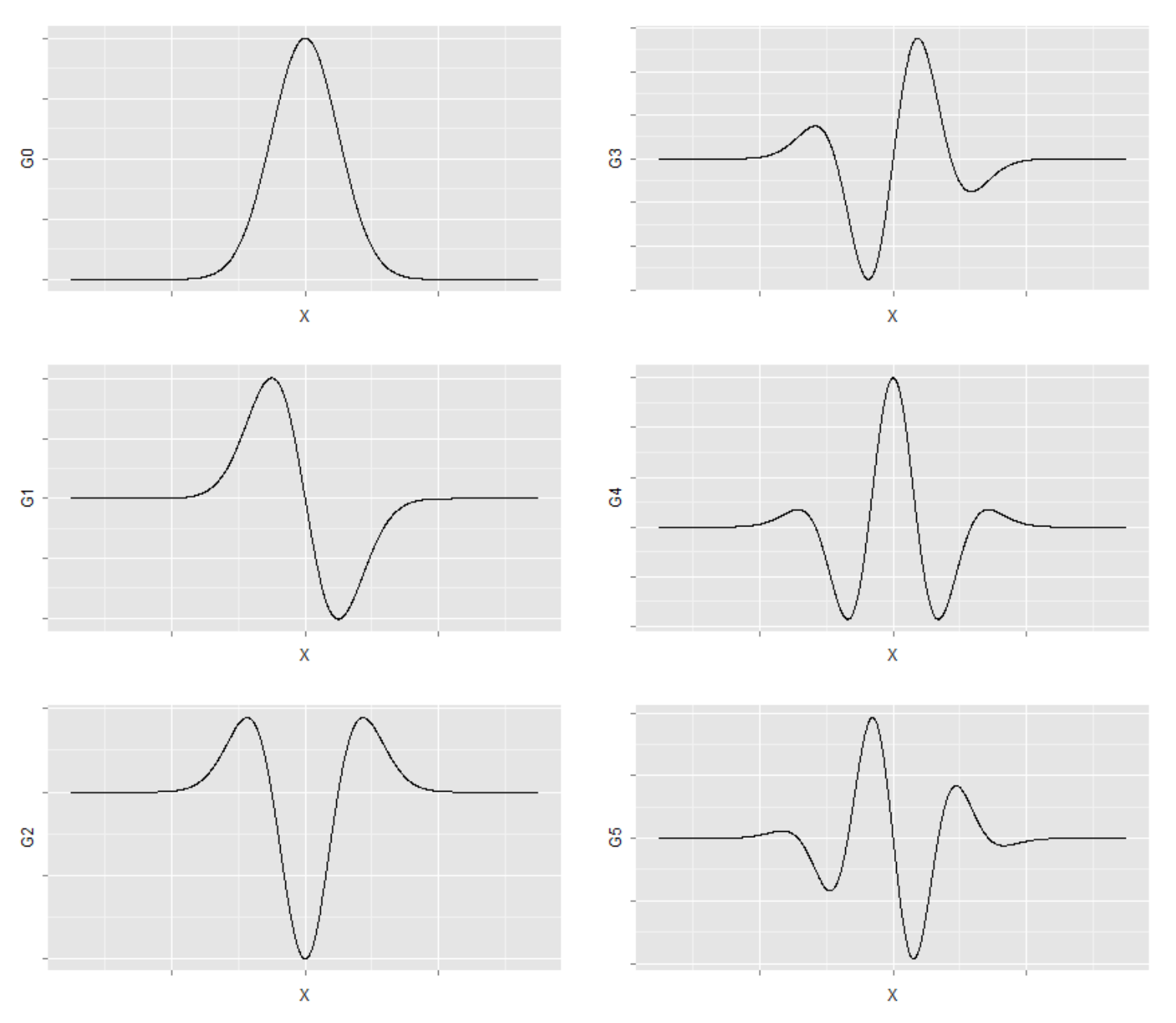}
			\caption{\textbf{Analyzing wavelets obtained from derivatives of the Gaussian function $\exp(-x^2/2)$.}
			\label{GaussWave}}
\end{figure}

Analyzing wavelets can be obtained from several ways. A classical one is to use the successive derivatives of the Gaussian function $\exp(-x^2/2)$. Indeed, the derivative of order $n$ is orthogonal to polynomials of order up to $n$ and of zero mean if $n$ is greater than 1. See Fig.~\ref{GaussWave} for a representation of these wavelets for order 0 to 5.

\begin{figure}
      \includegraphics[width=\columnwidth,keepaspectratio]{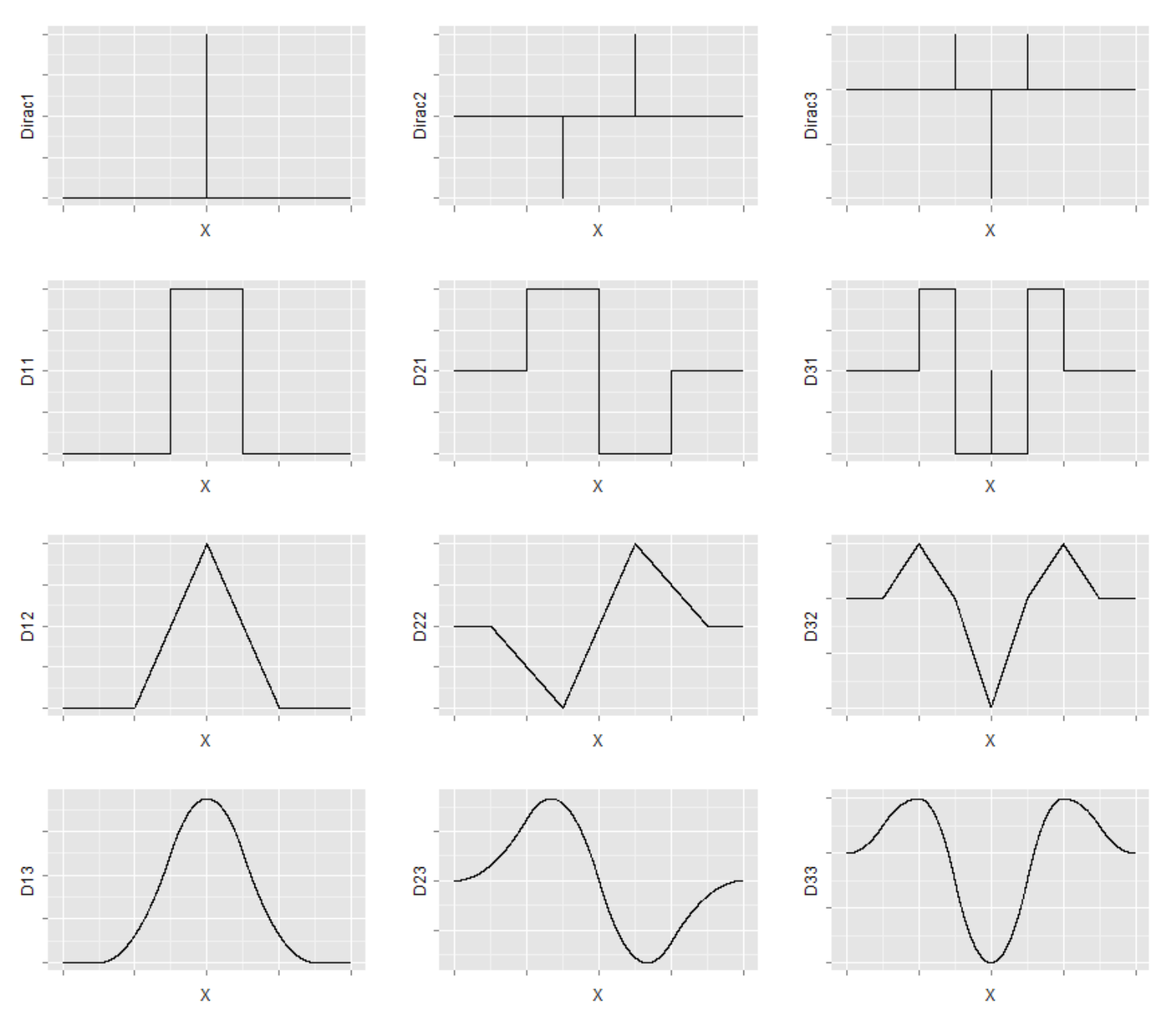}
			\caption{\textbf{Analyzing wavelets obtained from convolutions of the unit box over Dirac type distributions.}
			\label{AnaWave}}
\end{figure}

Another possible way is to process convolutions of the unit box over Dirac type distributions. On Fig.~\ref{AnaWave}, three successive convolutions of three variants of Dirac distributions are represented. The plot $Dij$ is obtained from the Dirac distribution Dirac$i$ by applying $j$ number of convolution. Note that only the last two Dirac distributions produce zero mean wavelets and that the unit box has been centered on 0 for aesthetic preferences.

WTMM can be easily extended to $n$ dimensions by considering the wavelets formed by the partial derivatives $\psi=(\psi_1,\cdots,\psi_n)$ of a function $\phi$ such as the Gaussian function $\exp(-\left|X\right|^2/2)$, where $X=(x_1,\cdots,x_n)$. The wavelet transform is then replaced by the higher dimensional version 
\begin{equation}
T_\psi[f](X_0,r):=\nabla T_\phi[f](X_0,r).
\end{equation}
More details on the two-dimensional case and examples are provided in \cite{AADMV}.

\begin{figure}[htp]
			\includegraphics[clip,width=0.9\columnwidth,keepaspectratio]{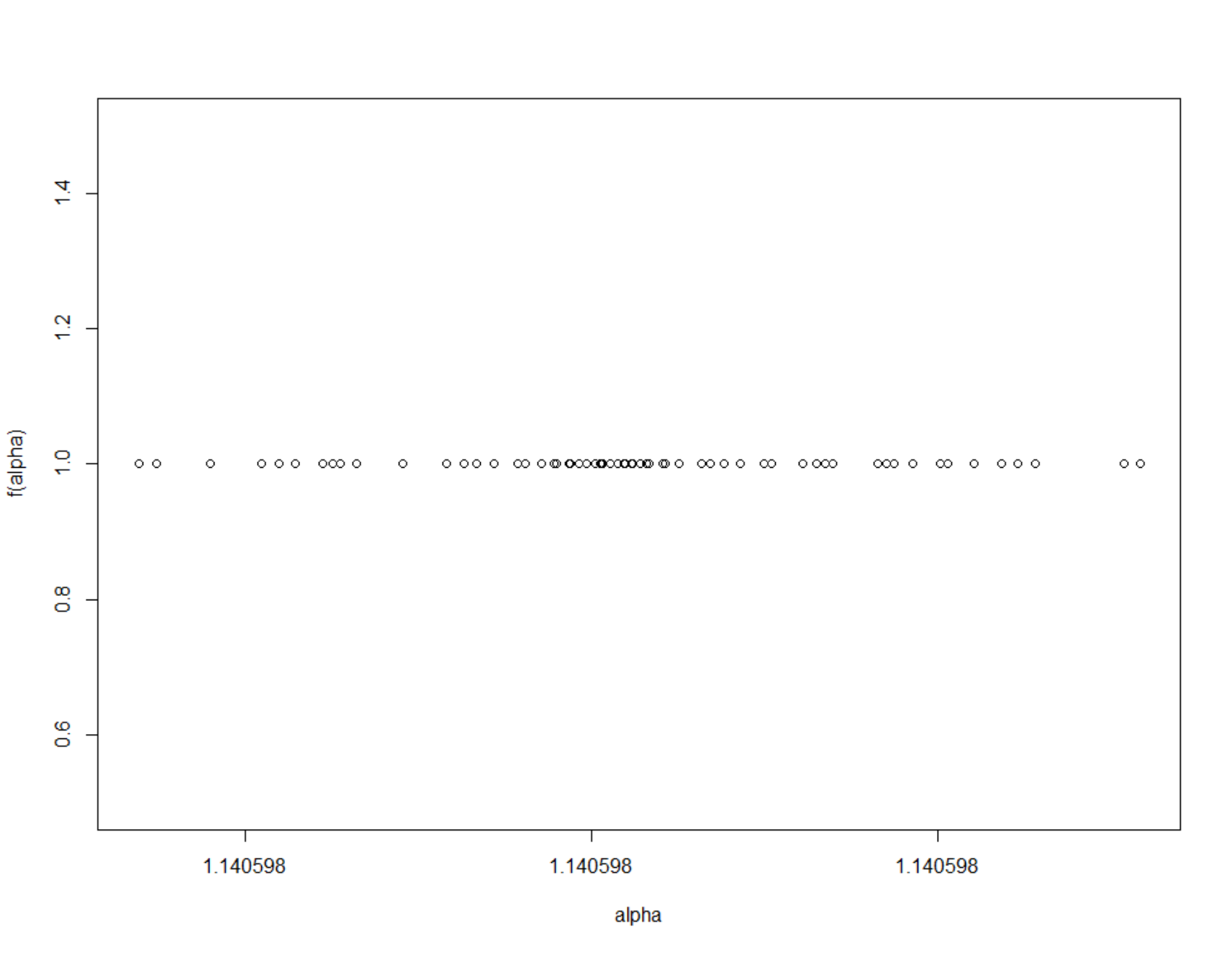}\\
			\includegraphics[clip,width=0.9\columnwidth,keepaspectratio]{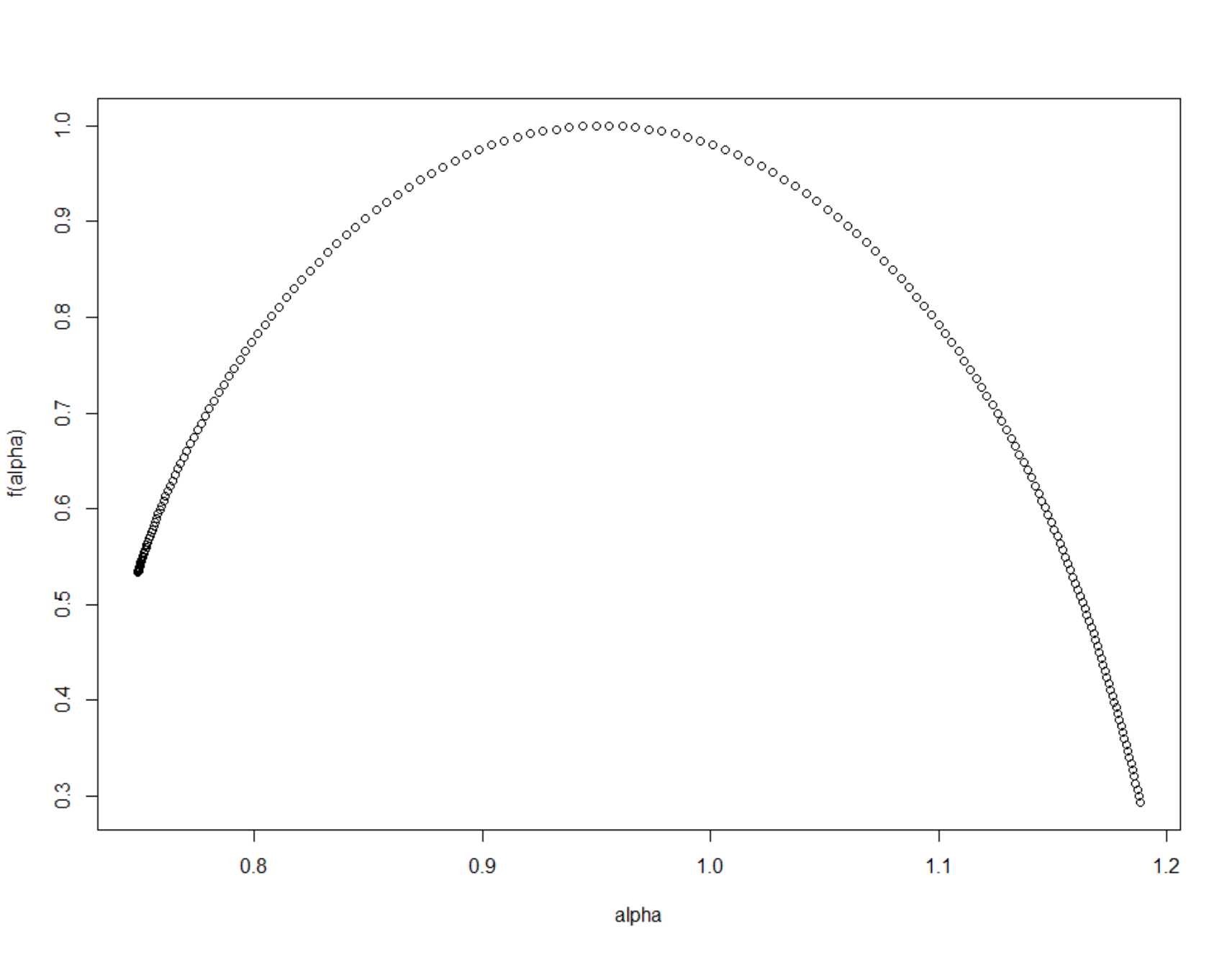}	
			\caption{\textbf{MDFA applied to binomial cascades.} On the top, a theoretical binomial cascade of parameter $p=0.6$ and on the bottom a particular realization of it.}
			\label{MDFA}
\end{figure}

In Fig.~\ref{MDFA}, MDFA is applied to the theoretical binomial cascade and to the random realization of it already used in section \ref{SpatData}. When the values of $s$ are chosen as powers of $2$ in the theoretical case, the $N_s$ intervals are only translated copies of themselves, resulting in a completely flat spectrum (on the top). MDFA is particularly well suited for data such as the random realization (on the bottom) and gives the closest results to what is expected from the mathematical study (see Fig.~\ref{BinomCascade}), with only a slight offset to the left of the range of $\alpha$ which is explained by the fact that the iteration process generating the cascade was stopped at a relatively low level of iterations.

\section{Interpretation and Limits}\label{interp}

The main elements of interpretation are presented in \ref{lpinterp} and links are established with some classical measures of heterogeneity such as Shannon's entropy. The type of information one can expect to gain from multifractal analysis is illustrated with the binomial cascade and results from a study on the multifractality of London's street network \cite{MMAB}. In \ref{limits}, limits of multifractal analysis are discussed, in particular the lack of consistency between methods.

\subsection{Link to usual measures and interpretation}\label{lpinterp}

Recall equation (\ref{BaseMoment}) and introduce the approximation that only the exponent $\tau(q)=\alpha(q) q-f(\alpha(q))$ is making a significant contribution to the value of $Z(q)$.
\begin{align}
Z(q) &\approx r^{\tau(q)}\int_\alpha \rho(\alpha)d\alpha \\
     &=r^{\tau(q)}.
\end{align}

Then, one defines the \emph{generalized dimension} as the family $\{D_q\}_q$, where
\begin{gather}
\forall q\neq1, \ D_q:=\lim_{r\rightarrow 0}\left[\frac{1}{q-1}\frac{\log Z(q)}{\log r}\right]=\frac{\tau(q)}{q-1},\label{Dq}\\
D_1:=\lim_{r\rightarrow 0}\frac{-\sum_ip_i\log p_i}{-\log(r)}.
\end{gather}
Three values of the generalized dimension are of particular interest. $D_0$ is the usual box-counting dimension of $D$ and therefore gives information on how much the data fills its physical support. $D_1$, referred to as the \emph{information dimension}, relates to Shannon's entropy and captures how even the data density is, with higher values of $D_1$ meaning a more uniform density. $D_2$ is the probability of pairs of independent events occurring in the same box and measures how scattered the data is, with increasing compactness for increasing values of $D_2$. It is similar to the correlation dimension quoted in \ref{monofrac}, see \cite{GP}.

\begin{figure}[htp]
			\includegraphics[clip,width=0.9\columnwidth,keepaspectratio]{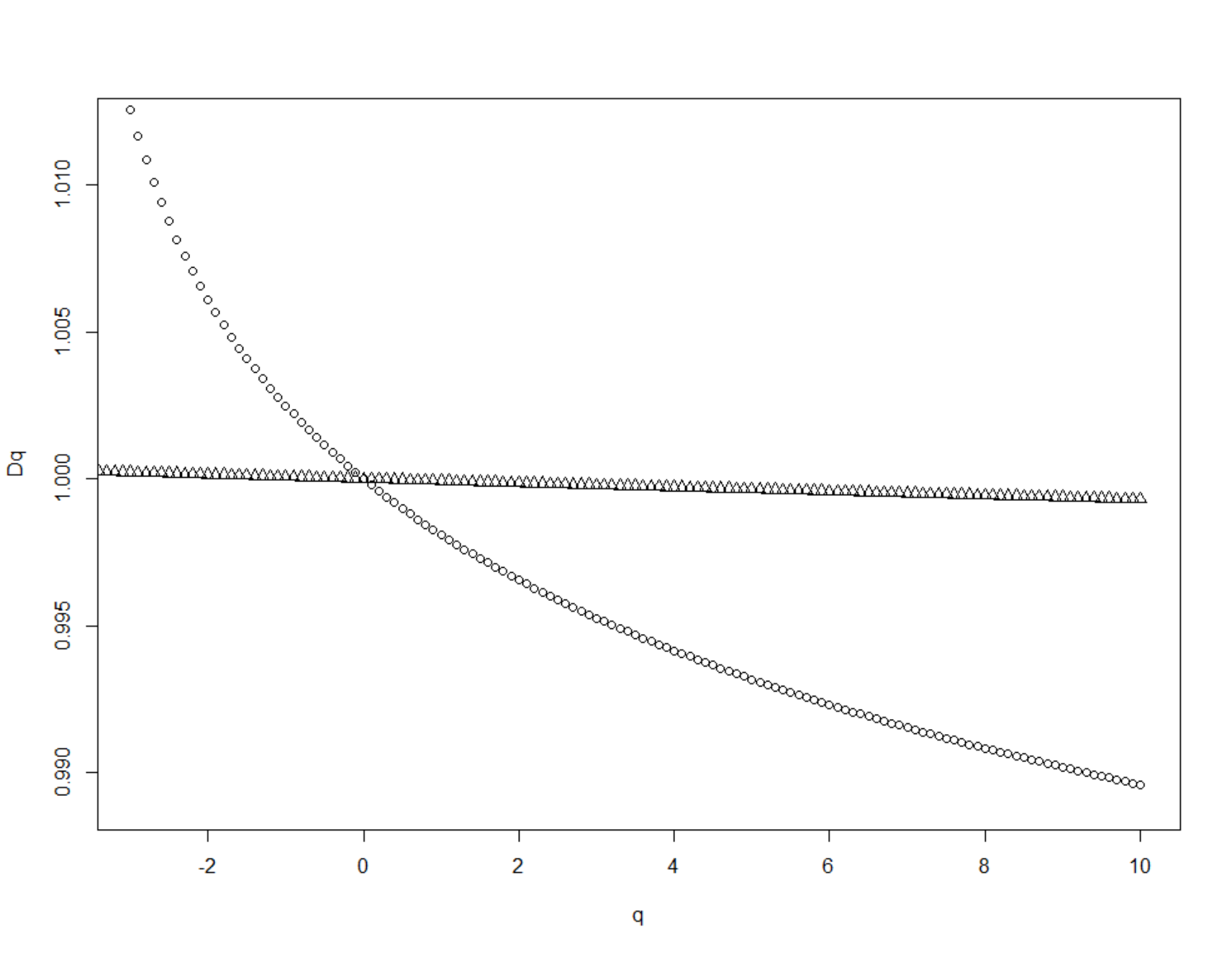}\\
			\includegraphics[clip,width=0.9\columnwidth,keepaspectratio]{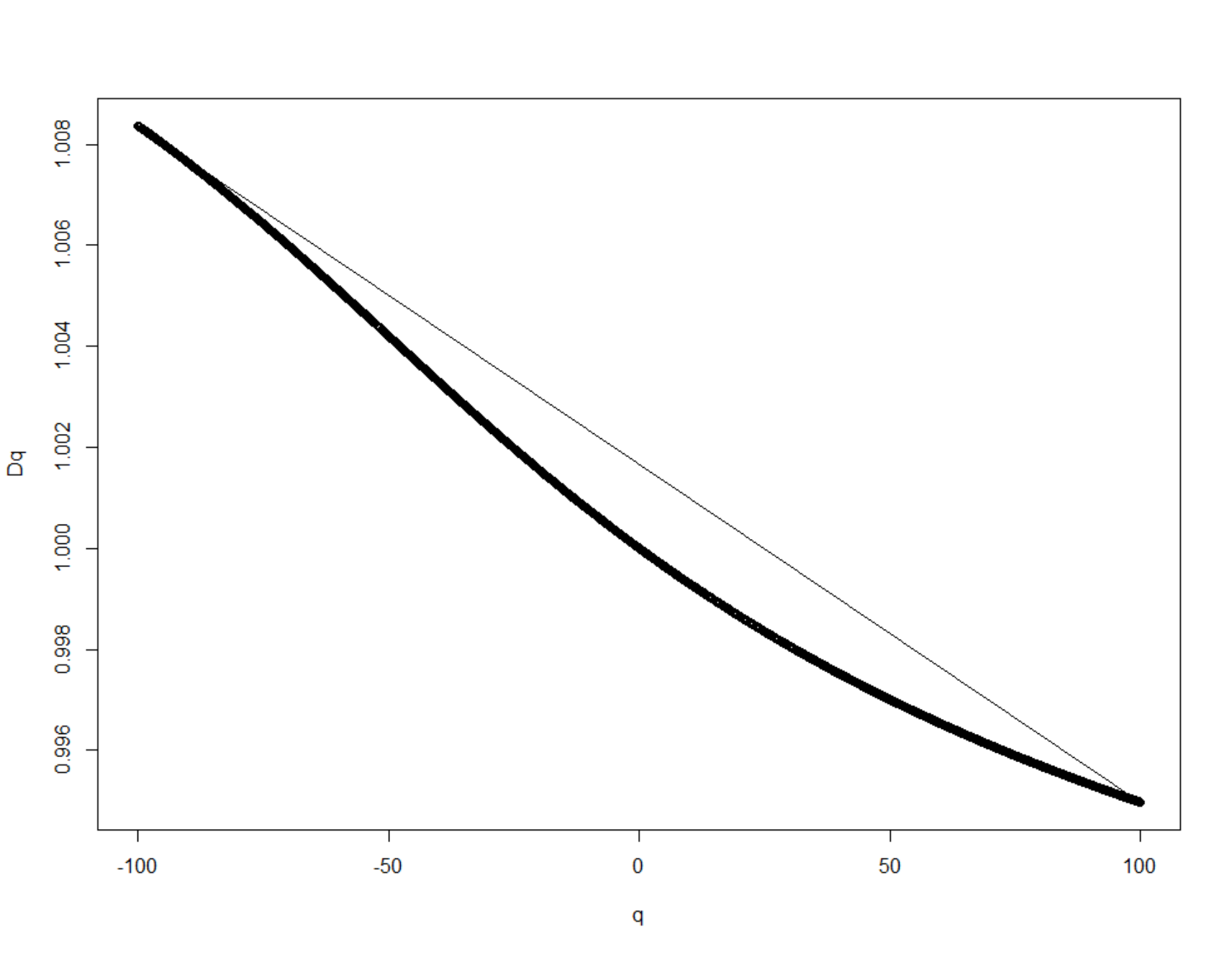}	
			\caption{\textbf{Comparison of two sets of $D_q$.} On the top, the standard moment method is applied to the theoretical random cascade of parameter $p=0.6$ (triangles) and to a particular realization of it (circles), showing weak multifractality in the theoretical case and stronger multifractality in the particular case. On the bottom, the $D_q$ curve of the theoretical case is zoomed in over a wider range of $q$ and compared to a straight line to show its slight curvature.}
			\label{Dqcomp}
\end{figure}

The full plot of $D_q$ versus $q$ is representative of the strength of the multifractality of $\mu$. The more constant the plot is, the weaker the multifractality is. See Fig.~\ref{Dqcomp}. It is also a good way to select the range of $q$ to study. Indeed, equation (\ref{Dq}) may lead to obtaining $D_q$ values greater than $d$ the dimension of $D$, in particular for negative values of $q$. As such a $D_q$ would loose its physical meaning, one may want to restrict the range of $q$ to ensure $D_q\leq d$. 

Another expression to obtain $D_q$ for $q\neq 1$ directly from $f(\alpha)$ and $\alpha$ is
\begin{equation}
D_q=\frac{f(\alpha)-q\alpha}{1-q}.
\end{equation}
It is mainly useful for the histogram method since $\tau(q)$ is never calculated when applying it. 

The limit of the generalized dimension is that it only gives global measurements of the whole data. In contrast, the multifractal spectrum gives one dimension for each set where the data scales similarly. In a sense, the variable $q$ selects different resolutions, with higher values of $q$ selecting a local scaling $\alpha(q)$ of lower order. The variable $f(\alpha(q))$ then gives the local fractal dimension at resolution $q$.

It is easy to see that the spectrum's peak is achieved for $q=0$, where $f(\alpha_0)=D_0$ is the fractal dimension of the physical support. Of particular relevance are therefore the asymmetries between the left and the right part of the spectrum. The spread of $\alpha$ indicates the variety of scaling present in the sample while the value of $f(\alpha)$ indicates the strength of the contribution of each $\alpha$.

For the binomial cascade, the range of $\alpha(q)$ is symmetric relatively to $q=0$ and the spectrum is quite ``round'', denoting a well balanced repartition of each scaling and its contribution. A more interesting example is given by the evolution of London's street network.

\begin{figure}[htp]
			\includegraphics[clip,width=0.9\columnwidth,keepaspectratio]{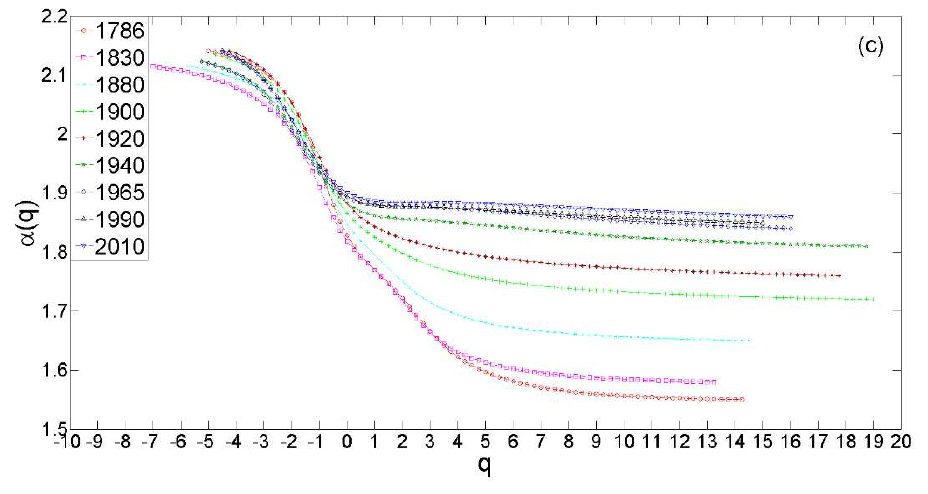}\\
			\includegraphics[clip,width=0.9\columnwidth,keepaspectratio]{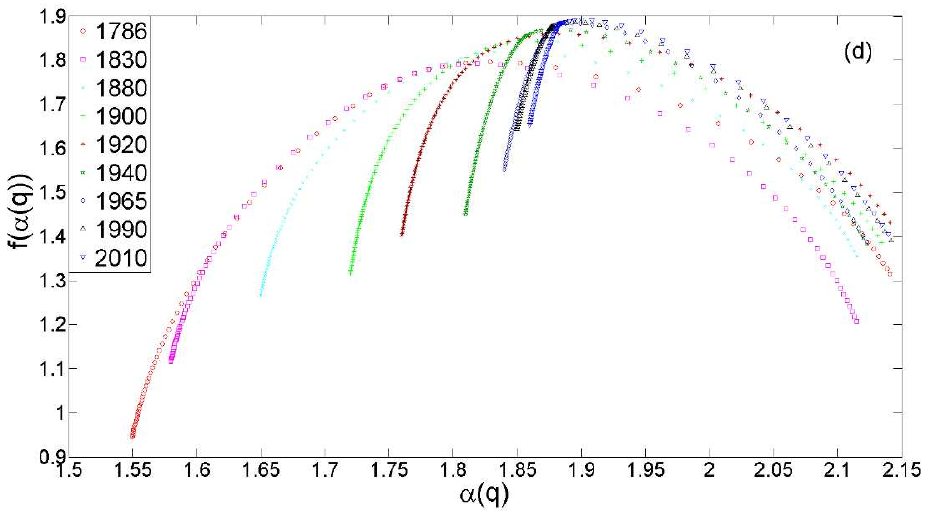}	
			\caption{\textbf{London's street network multifractality} \textit{(taken from \cite{MMAB}).} On the top, the curve $\alpha$ versus $q$ shows the variety of zones with low probabilities of intersection not evolving much over the years (left part of the curve), while disappearing in zones corresponding to higher probabilities (right part of the curve). On the bottom, the spectrum indicates an evolution that favours more and more areas with an increasingly lower density of intersections while presenting less and less variety.}
			\label{Rob}
\end{figure}

Fig.~\ref{Rob} presents a clear picture of the structural differences that the London's street network has experimented in the last 200 years. The strong differences in value and shape between the first and last years is an evidence of how this street network evolved to lose its multifractal nature, and has become more homogeneous in terms of intersection density across the whole city.

Firstly, the singularity exponent $\alpha(q)$ accounts for the balance between areas with more/less street intersections (Fig.~\ref{Rob}, top). In 1786, the $\alpha(q)$ values for positive $q$ are relatively low compared with the $α(q)$ values for negative $q$, a situation that confirms that the number of areas with major intersection densities are not that common. As we move forward in time, these differences are less and less evident, until 2010, where basically the same density can be found across the whole network.

The multifractal spectrum curves (Fig.~\ref{Rob}, bottom) represent the distribution of intersection densities between the different regions. The left part of each curve (related with positive $q$, i.e., with the denser areas) became less and less wide as we move forward in time, while the right section remains stable for all nine networks. This is a clear indicator of how most of the network is evolving to become more similar through its different areas.

\begin{figure}
      \includegraphics[width=0.9\columnwidth,keepaspectratio]{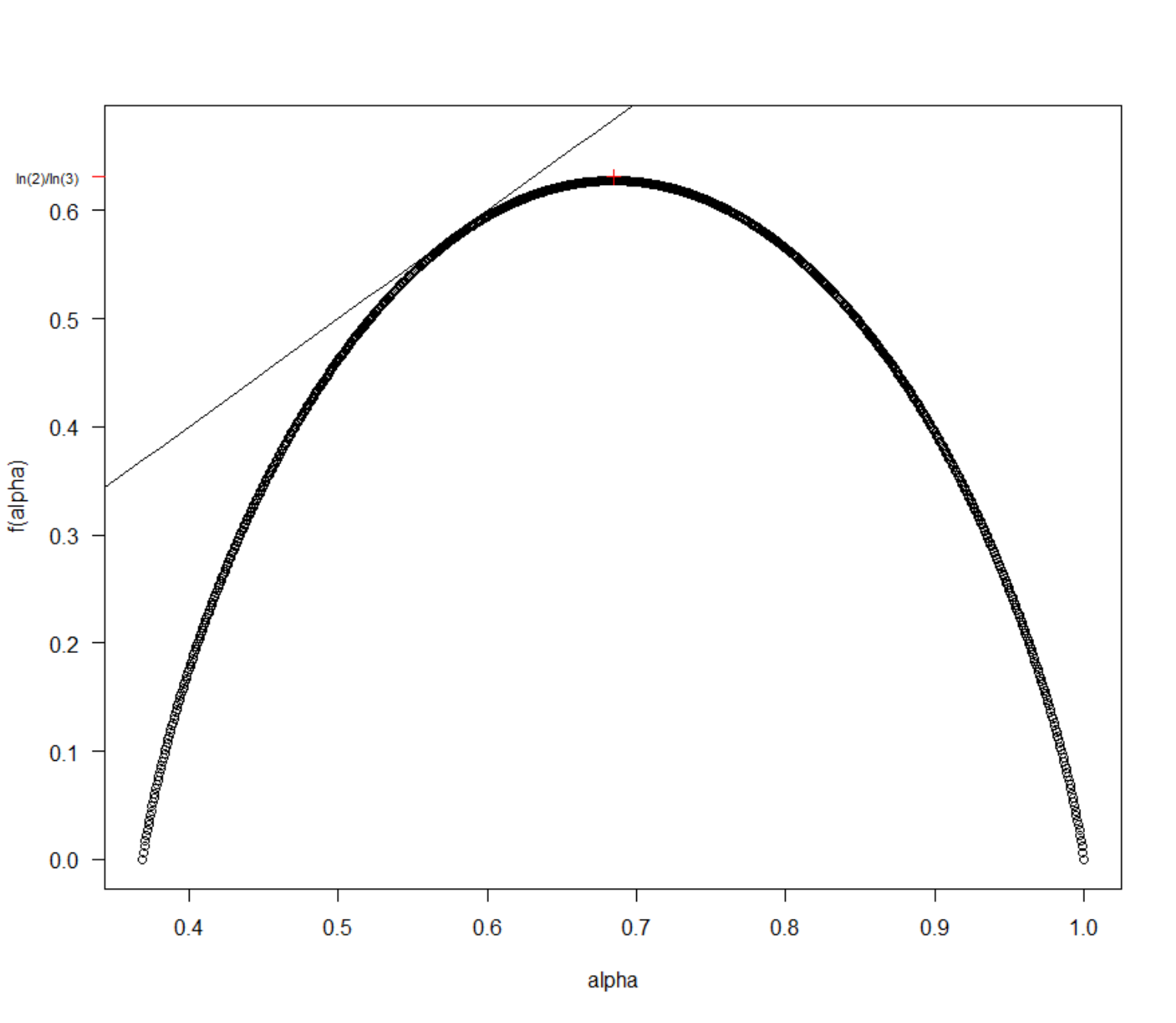}
			\caption{\textbf{Tangency with the diagonal.} On the spectrum obtained for the middle third Cantor set, the diagonal $f(\alpha(q))=\alpha(q)$ is tangent to the spectrum on the point where $q=1$.}
			\label{Diag}
\end{figure}

One last property of the curve $f(\alpha)$ versus $\alpha$ can be used to test the correctness of the result: since $\mu$ is finite, equation (\ref{tauq}) implies that $\tau(1)=0$ and, consequently, that $f(\alpha(1))=\alpha(1)$. Moreover $df(\alpha)/d\alpha=q=1$, implying that the spectrum lies below the diagonal and touches it exactly in the point corresponding to $q=1$. See Fig.~\ref{Diag} for the middle third Cantor set.

\subsection{Limits}\label{limits}

Using a multifractal study written by someone else can be a perilous initiative. The definition of what is called here $\M$ is often left implicit and some authors restrict the analysis to either the multifractal spectra or the generalized dimension, or even to the three values $D_0$, $D_1$ and $D_2$ only. Furthermore, most methods are equivalent for rigorously abstract self-similar sets and measures, but practical estimations often create an overshooting that is hard to measure. It is unfortunately a necessity to tailor the experimental protocol to the type and amount of data studied. One should therefore be extremely careful when comparing analyses made under different circumstances.

Results of isolated multifractal studies should be seen as giving ``trends'' or describing an evolution. In \cite{MMAB} for example, the loss of multifractality in London's street pattern through time is striking. The study provides rather precise information on which part of the spectrum is collapsing and from what point in time it becomes noticeable, only because it is consistently applied over sufficiently similar datasets of London's street network. 

Another main issue when computing the spectra is controlling the error. To the theoretical error inherent to the chosen method one has to add the practical concessions for computational efficiency, and the measurement error that inevitably arises from the large amount of data required. The most common sources of error are the edge effects and the linear fit that is generally used to deduce $\tau(q)$ from the slopes of $Z(q)$ against $r$ in log-log plots. The value $\tau(q)$ is quite sensitive and its computation often has to be automated and operated over only few values of $r$.

Choosing an adequate range of $q$ is critical for moment based methods (including MDFA). In theory, $q$ should go from $-\infty$ to $+\infty$, which is not possible in practice. For negative values of $q$, one may find values of $D_q$ greater than the dimension of the support of the measure, indicating that those values of $q$ should be discarded. Furthermore, values for which the linear fitting of $Z(q)$ versus $r$ is not obtained with a sufficiently good level of confidence should also be discarded. For example, Lee and Stanley found in \cite{LS} that for diffusion-limited aggregation, a phase transition occurs. Below some critical value of $q$, the function $Z(q)$ does not scale as a power law and forcing a linear fit for these values would create important discrepancies.

As evidenced through the example of the binomial cascades, the smallest resolution allowed by the studied dataset may play an important role in the results' accuracy. Multifractality aims at studying measures at an infinitely small scale and a scale as small as $10^{-6}$ might not be small enough to be considered ``infinitely'' small. In particular, when comparing two similar objects for which the measure cannot be evaluated with the same level of precision, one should use the largest minimal resolution of the two as the starting one for both.

It should be noted that predicting beforehand the shape of the spectrum is often difficult. There exist exactly self-similar nonrandom measures for which half of the spectrum is not defined \cite{MEH}, proving that unusual shapes may be normal.	Another remark is that if $\mu_1$ and $\mu_2$ are finite measures on $\R^n$ with disjoint supports, then it can be proven that 
\begin{equation}
f_\H^{\mu_1+\mu_2}(\alpha)=\max\left\{f_\H^{\mu_1}(\alpha),f_\H^{\mu_2}(\alpha)\right\},
\end{equation}
where $f_\H^\mu$ is the fine spectrum of measure $\mu$ as defined in \ref{mathdef}. In particular, $f_\H(\cdot)$ does not need to be concave.

\section{Conclusions}

Four methods were presented in the multifractal context. They are the moment method, the histogram method, MDFA and WTMM. Each is particularly well suited for a particular type of data: spatial data for the moment method and also for the histogram method provided the range of scaling is large enough, one-dimensional time series for MDFA and WTMM, with a convenient possibility to extend to higher dimensions for the last one.

Through the study of binomial cascades, it was found that the MDFA method gave the best results in addition to the possibility to detect the source of multifractality by applying the shuffling procedure. The moment method proved both simple and reliable with many variants giving it great adaptability to different contexts.

It was finally evidenced that multifractality can provide useful information about both the local and the global complexity and inhomogeneity of a phenomenon. In addition, it can be an effective comparative tool for measures and spaces that are too complex for classical geometry provided the methodology used is consistent. It was emphasized that one should focus on trends and evolutive aspects in a phenomenon rather than expect to obtain a foolproof numerical result.

\appendix

\section*{Appendix: R Codes}

The codes implemented in R for the first three methods described in this paper will be available at the following link.

\bibliography{Multifractals_article_bib}

\end{document}